\documentclass[manuscript, screen, authorversion]{acmart}

\AtBeginDocument{%
  }

\copyrightyear{2025}
\acmYear{2025}
\setcopyright{acmlicensed}
\acmConference[AutomotiveUI '25]{17th International Conference on Automotive User Interfaces and Interactive Vehicular Applications}{September 21--25, 2025}{Brisbane, QLD, Australia} 
\acmBooktitle{17th International Conference on Automotive User Interfaces and Interactive Vehicular Applications (AutomotiveUI '25), September 21--25, 2025, Brisbane, QLD, Australia}
\acmDOI{10.1145/3744333.3747822} 
\acmISBN{979-8-4007-2013-0/2025/09}

\acmSubmissionID{1039}

\usepackage{acmart-taps}
\usepackage[T1]{fontenc}
\usepackage{xcolor}
\usepackage{xspace}
\usepackage[capitalise]{cleveref}
\usepackage{makecell}
\usepackage{booktabs,subcaption,amsfonts,dcolumn} 

\newcommand{\ie}{\emph{i.e.}\xspace}
\newcommand{\eg}{\emph{e.g.}\xspace}

\definecolor{Green}{RGB}{0,128,0}
\newcommand{\LevelOne}{\textcolor{Green}{Level~1}\xspace}
\newcommand{\LevelTwo}{\textcolor{orange}{Level~2}\xspace}
\newcommand{\LevelThree}{\textcolor{red}{Level~3}\xspace}

\begin{document}

\title[Effects of Cognitive Distraction and Driving Environment Complexity on Adaptive Cruise Control Use]{Effects of Cognitive Distraction and Driving Environment Complexity on Adaptive Cruise Control Use and Its Impact on Driving Performance: A Simulator Study}

\author{Ana{\"i}s Halin}
\email{anais.halin@uliege.be}
\orcid{0000-0003-3743-2969}
\affiliation{%
  \institution{University of Liège}
  \department{Department of Electrical Engineering and Computer Science}
  \city{Liège}
  \country{Belgium}
}

\author{Marc Van Droogenbroeck}
\email{m.vandroogenbroeck@uliege.be}
\orcid{0000-0001-6260-6487}
\affiliation{%
  \institution{University of Liège}
  \department{Department of Electrical Engineering and Computer Science}
  \city{Liège}
  \country{Belgium}
}

\author{Christel Devue}
\email{cdevue@uliege.be}
\orcid{0000-0001-7349-226X}
\affiliation{%
  \institution{University of Liège}
  \department{Department of Psychology}
  \city{Liège}
  \country{Belgium}
}


\begin{abstract}
In this simulator study, we adopt a human-centered approach to explore whether and how drivers' cognitive state and driving environment complexity influence reliance on driving automation features. Besides, we examine whether such reliance affects driving performance.
Participants operated a vehicle equipped with adaptive cruise control (ACC) in a simulator across six predefined driving scenarios varying in traffic conditions while either performing a cognitively demanding task (\ie, responding to mental calculations) or not. 
Throughout the experiment, participants had to respect speed limits and were free to activate or deactivate ACC.
In complex driving environments, we found that the overall ACC engagement time was lower compared to less complex driving environments.
We observed no significant effect of cognitive load on ACC use. 
Furthermore, while ACC use had no effect on the number of lane changes, it impacted the speed limits compliance and improved lateral control.
\end{abstract}



\maketitle


\section{Introduction}
The Society of Automotive Engineers (SAE)~\cite{Sae2021Taxonomy} defines six levels of driving automation, ranging from no automation (SAE Level 0) to full automation (SAE Level 5). 
Vehicles currently on the market are mainly SAE Level 2 (partial driving automation), such as Tesla Autopilot, BMW Extended Traffic Jam Assistant, Ford Blue Cruise, and many more~\cite{Sever2024Automated}.
In 2022, Mercedes-Benz's system called \linebreak[3] ``Drive Pilot'' was the world’s first fully certified SAE Level 3 driving system (conditional driving automation).
Initially restricted to geo-fenced areas on the German highway at speeds up to $60$~km/h, an updated version available in 2025 extends its use to flowing traffic up to $95$~km/h under certain conditions on the entire German highway network.
Despite these advancements, fully autonomous SAE Level 5 vehicles remain a distant reality, meaning that driving will remain a collaboration between humans and autonomous driving systems in the foreseeable future~\cite{Xing2021Toward}. It is therefore essential to ensure that this collaboration is as effective as possible in terms of safety, driver comfort, and overall performance. 

These SAE levels provide a useful classification for regulatory bodies and automotive manufacturers, enabling legal and technical classification of vehicles based on their level of automation. Yet, they do not fully capture how automation operates in real-world driving and offer limited practical value for drivers. Indeed, automation capabilities do not remain static throughout a drive, and specific features may switch off in some situations. For instance, in a SAE Level 3 vehicle, control may shift between the driver and the automation system. Moreover, even when the driver remains responsible for driving (SAE Levels 0–2), they may activate or deactivate specific assistance features (\eg, adaptive cruise control (ACC) or lane-keeping assistance) at will, dynamically adjusting the level of automation.

For safe and effective human-automation interaction, drivers must accurately understand their moment-to-moment responsibilities, and the capabilities and limitations of the automation in use~\cite{Novakazi2021Levels, Blomacher2020TheEvolution, Buchner2024Exploring}. 
However, research has shown that drivers often lack awareness of automation limitations~\cite{Beggiato2013TheEvolution}, which can result in misuse, disuse, or overreliance. The phenomenon of \emph{autonowashing}~\cite{Dixon2020Autonowashing}, where marketing exaggerates automation capabilities, leading to misinterpretations of system reliability, further compounds this issue.
Given these challenges, we argue that a \emph{dynamic operational level of automation}, which fluctuates over time as driving automation features are engaged or disengaged, is more relevant to day-to-day driver safety than the static SAE Level. 
Moreover, this operational level of automation could be dynamically adapted based on both the driver state and the driving environment, a concept known as \emph{adaptive automation}~\cite{Cabrall2018Adaptive}.

The primary motivation for integrating automation into vehicles is to enhance safety and comfort for drivers and passengers. Automation may contribute to this objective by relieving the driver from parts of the driving task. Yet, excessive automation may have unintended consequences, leading to reduced situation awareness and disengagement from the driving task, which in turn increase risks~\cite{DeWinter2014Effects, Bai2025Awakening}.
\citet{Young2023ToAutomate} advocate for the `cliff-edge' principle, which consists in restraining automation capabilities until it can fully and reliably take over the task at hand, and favor human-centered support rather than technology-centered automated replacement.
Given that drivers will still be involved in the control loop for some time, they should be actively involved rather than passively supervising.
Similarly, \citet{Endsley2016From} emphasize that automation should only be used where necessary and at the lowest possible level.

Adaptive automation offers a potential solution to these concerns by dynamically adjusting the operational automation level in response to both the driving environment and the driver state. This approach seeks to balance automation and driver involvement, activating automation only when beneficial and without compromising safety. Ideally, adaptive automation should help maintain optimal driver involvement, preventing both overreliance and underuse. Furthermore, in situations where a high level of automation is temporarily feasible, it should allow for a smooth transition back to driver control, providing sufficient time for drivers to regain situation awareness.
A key challenge is determining when and how automation should take over versus when it should promote driver involvement. 

In this study, we investigate how driver state and driving environment complexity influence the use of driving automation features and how automation, in turn, impacts driving performance. Our findings contribute to the broader discussion on adaptive automation by providing insights into how automation can be designed to support rather than replace human drivers, ultimately enhancing safety.
Specifically, we conducted a driving simulator study in which participants could freely activate or deactivate ACC while driving under different conditions across six scenarios. These scenarios varied along two dimensions: driver state, with two levels of cognitive distraction (\ie, presence/absence of a secondary mental calculation task), and driving environment complexity, with three levels (\ie, increasing traffic density and addition of road constructions restricting the number of traffic lanes).


\section{Related Work}
\subsection{Driving Environment Complexity} 
The complexity of the driving environment is influenced by many factors, such as weather conditions, road types, or traffic density. For example, urban driving is considered more complex than rural driving~\cite{Buchner2024Exploring}.

The driving environment complexity influences the use of driving automation features or systems, either because of system limitations~\cite{Sae2021Taxonomy}, regulations~\cite{Sever2024Automated}, or driver preferences~\cite{Wu2015Drivers,Gershon2021Driver}. For instance, the use of driving automation features is usually recommended only in good weather or other beneficial driving conditions~\cite{Sever2024Automated}, and drivers tend to not activate ACC on city streets that have traffic lights~\cite{Wu2015Drivers}.

The impact of driving environment complexity is particularly studied for SAE Level 3 vehicles, where the driving automation system reaches its limitations and issues a take-over request (TOR), requiring the driver to quickly resume control. Research indicates that traffic density influences both take-over time and quality~\cite{Radlmayr2014How, Gold2016Taking, Du2020Evaluating}, whereas weather conditions, such as fog-induced low visibility, do not significantly affect take-over performance~\cite{Lin2024How}. Cooperative perception of the driving environment~\cite{Naujoks2016AHumanMachine} helps reduce risks associated with TORs. By combining vehicle-localized environment perception (\eg, via cameras and LiDAR) with information from other road users or infrastructure, it allows TORs to be issued earlier, giving the driver more time to get back into the loop while the system is still active. 

\subsection{Driver State} 
Driver state is commonly categorized into five (sub-)states: drowsiness, mental workload, distraction, emotions, and under influence. Distraction is further divided into manual, visual, auditory, and cognitive distraction~\cite{Halin2021Survey}. 

Maintaining an optimal mental workload is important for effective task performance, as both underload and overload are associated with decreased driving performance~\cite{Yerkes1908Relation, Young2002Attention, Coughlin2011Monitoring}. 
The impact of driving automation on driver workload remains a topic of debate. Some studies suggest that automation reduces workload~\cite{Stanton2005Driver}, while \citet{Vasta2025Effect} conducted a meta-analysis that found no significant reduction in mental workload between manual driving (SAE Level 0) and partially automated driving (SAE Level 2). They argue that workload reduction may only occur in drivers with automation experience or after extended use during long drives.
However, it is widely accepted that driving automation reduces stress~\cite{Stanton2005Driver} while also lowering situation awareness~\cite{Stanton2005Driver, DeWinter2014Effects} and increasing drowsiness~\cite{Vogelpohl2019Asleep, Ahlstrom2021Effects}, distraction, and engagement in non-driving-related tasks (NDRTs)~\cite{Vasta2025Effect}.
While much research has examined how automation affects driver state, less attention has been given to how driver state influences the use of automation.

Studies indicate that factors contributing to the complexity of the driving environment have little to no effect on driver state, while others do. For instance, \citet{Ahlstrom2018TheEffect} reported that light conditions (daylight versus darkness) had a small impact on sleepiness-related measures (\eg, subjective sleepiness, lateral position, speed, or blink durations). However, a more recent study~\cite{Meyerson2024Effects} found no significant effect of light conditions on driver sleepiness. Similarly, road environment (rural versus suburban) was shown to have little influence on driver sleepiness~\cite{Ahlstrom2018Effects}.
In contrast, other factors appear to have stronger effects. \citet{Stanton2005Driver} found that high traffic density significantly increases both driver mental workload and stress, highlighting the importance of considering these interactions when designing adaptive automation systems.
Similarly, \citet{Park2022TheImpact} found that environmental factors (such as visual complexity or number of objects in the visual scene)  significantly impact driver situation awareness.

\subsection{Driving automation}
Driving automation systems consist of a collection of features that can perform all or part of the dynamic driving task, depending on the level of automation~\cite{Sae2021Taxonomy}. 
As such systems become increasingly integrated into production vehicles, research has expanded to investigate how drivers interact with these features (for example, during transitions of control~\cite{Gershon2021Driver,Herzberger2022Confidence,Herzberger2024Cooperation}), how cooperation between driver and automation can be supported~\cite{Walch2019Driving,Griffiths2005Sharing}, and how human–machine interfaces can enhance these interactions~\cite{Dixon2025Explaining}.

For instance, \citet{Gershon2021Driver} analyzed how drivers leverage automation in real-world driving, focusing on SAE levels 0, 1, and 2, and identified driver and system-initiated transfers of control. They found that drivers frequently initiated transitions between automation levels, and that these were not necessarily related to immediate risk mitigation, but often due to a mismatch between system capabilities and driver expectations or preferences. While this work sheds light on drivers’ reasons for engaging or disengaging automation (\eg, disengaging ACC), it does not address how these decisions are modulated by the driving environment complexity or the driver state. 

The influence of driving context on automation use has been highlighted by \citet{Orlovska2020Effects}, who demonstrated that the relationship between driver, system, and environment is highly interdependent. However, their study also did not incorporate the impact of driver state, such as cognitive distraction, on automation use.


\section{Research Questions}

In this study, we examined the relationships between driver state, driving environment complexity, and driving automation. Specifically, we aimed at answering the following four research questions (RQs):
\begin{itemize}
    \item[\textbf{RQ1}] How does driving environment complexity influence ACC use?
    \item[\textbf{RQ2}] How does cognitive distraction influence ACC use? 
    \item[\textbf{RQ3}] How does ACC use influence driving performance in different driving environments?
    \item[\textbf{RQ4}] How does ACC use influence driving performance in the presence or absence of a cognitive distraction?
\end{itemize}


\section{User Study}
To analyze the relationships between driver state, driving environment complexity, and driving automation, we conducted a within-subjects driving simulator experiment. 
Participants drove the same route six times, experiencing three levels of environmental complexity, once with a cognitively demanding secondary task and once without (two levels of cognitive distraction).
Participants were instructed to respect traffic laws, including speed limits, and were free to activate or deactivate the ACC at any time. 

We investigated whether the activation or deactivation of driving automation features depended on the driver state and/or on the driving environment complexity. We also examined whether ACC use affected driving performance. 

\subsection{Participants}
A total of $31$ participants were recruited. However, $2$ participants did not complete the experiment due to simulator sickness, leaving a final sample of $N=29$ participants ($22$ male, $7$ female; mean age $= 31.59$, SD $= 10.68$ years). Data from the $2$ withdrawn participants were entirely discarded for the analysis.
Participants were recruited via posters, word-to-mouth or social media (\eg, Facebook, LinkedIn).
All participants had a valid driver’s license, on average, for $11.62$ (SD $= 11.18$) years.  
The study was approved by the ethics committee of the Faculty of Psychology, Logopedics and Educational Sciences of the University of Liege under the reference 2223-081. 
Participants all provided informed and signed consent before taking part in the experiment.

\subsection{Apparatus and Materials}
\subsubsection{Driving Simulator}
\begin{figure*}
    \centering
    \includegraphics[width=.60\linewidth]{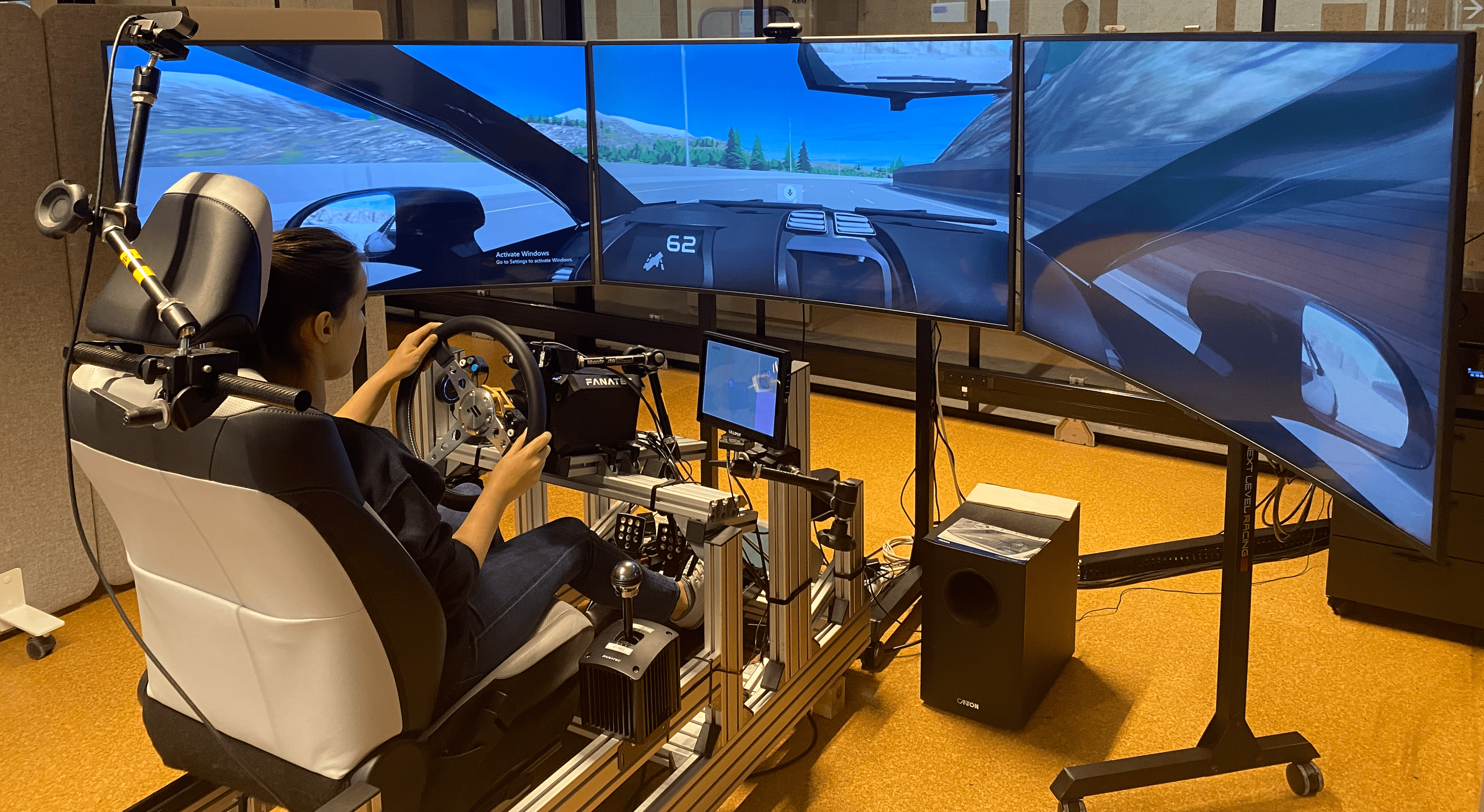}
    \Description{This image shows a view of the simulator with a driver and three screens in front of him/her.}
    \hfill
    \includegraphics[width=.35\linewidth]{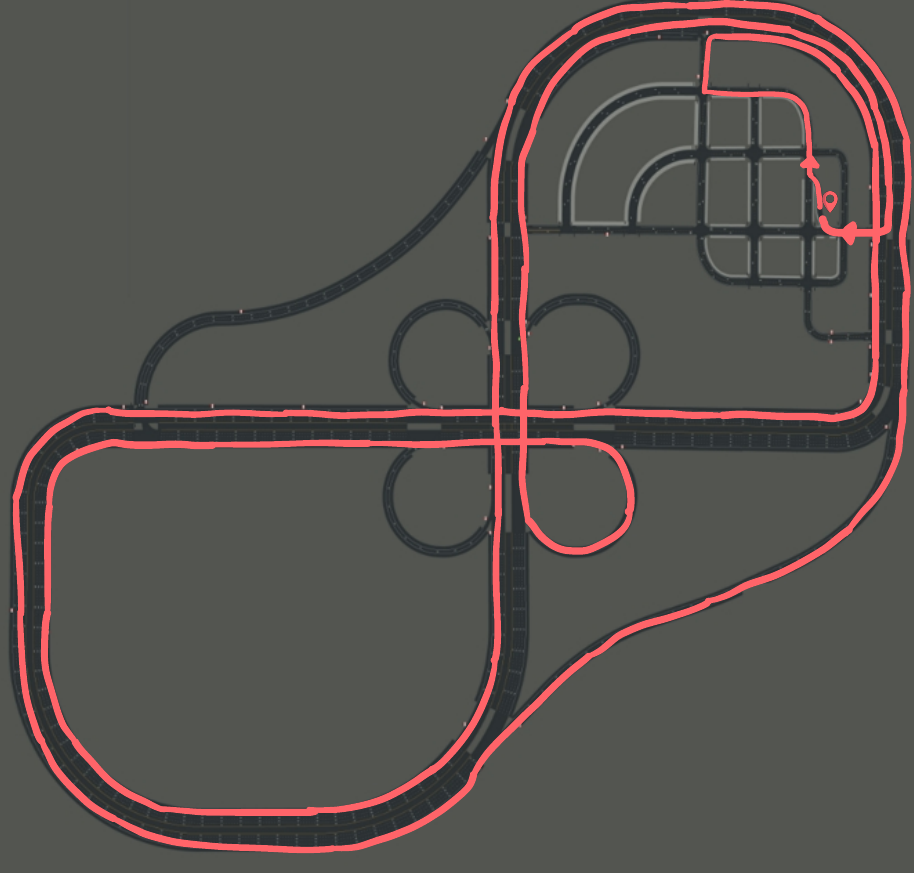}
    \Description{This figure shows the top map of the vehicle circuit named ``Town 4''.}
    \caption{Illustration of the experimental setup. A participant is seated in the driving simulator (left) and performing six scenarios on the same standardized CARLA map named ``Town 4'' (right). The route consisted of a segment through a small town with pedestrians and traffic lights, followed by an infinite loop highway section. The person depicted provided permission.}
    \label{fig:simulator}
\end{figure*}

The experiment was conducted in a driving simulator developed by AISIN Europe, running a customized version of CARLA~\cite{Dosovitskiy2017CARLA}, which is an open-source simulation environment based on Unreal Engine. The setup included three large $50$-inch curved screens, an adjustable car seat, and a Fanatec system comprising a steering wheel, gear shifter, and pedals. 
The driving simulator featured an automatic transmission vehicle. Turn signals (blinkers) were controlled via buttons on the steering wheel. The vehicle was equipped with an ACC system, which not only adapted to the speed of the preceding vehicle but also automatically slowed down in curves. The ACC implementation was inspired by~\cite{Watzenig2017Automated} and based on a carrot-chasing algorithm~\cite{Micaelli1993Trajectory,Sujit2014Unmanned}. Participants were able to engage ACC at their current speed and adjust it in increments or decrements of 5 km/h using dedicated buttons on the steering wheel. ACC could only be disengaged via these buttons; pressing the brake pedal neither deactivated ACC nor initiated braking while ACC was engaged. 
The ACC system did not automatically adapt speed based on recognized speed limit signs, as no signs were present. Instead, speed limits were communicated to participants via a voice agent. Consequently, when using ACC, participants were responsible for manually adjusting the set speed to comply with speed limits (\eg, in road construction zones). Additionally, the vehicle was equipped with a blind spot detection system.
Participants also had a small screen on their right to mimic a navigation system.
Experimental sessions were recorded using two 4K cameras capturing front and back views, a high-resolution infrared camera positioned behind the wheel as part of a Driver Monitoring System (DMS), and a microphone for audio recording. 
\Cref{fig:simulator} shows the experimental setup.
A custom software suite, developed by AISIN Europe, was used to design the test scenarios, execute the simulations, and log experimental data.

\subsubsection{Pre-Test and Feedback Questionnaires} 
The pre-test questionnaire aimed at collecting demographic and study-specific data. 
To gain insights on participants' experience and habits, we asked them to rate their familiarity with, use of, and attitude toward ACC, driving automation features and systems on five-point Likert scales.
In addition, after each scenario, participants completed a feedback questionnaire, collecting, \eg, information about the reasons for engaging or not engaging the ACC. 
Both questionnaires are available at \url{https://osf.io/t3ug7}. 

\subsection{Tasks}
\subsubsection{Driving Task}
Participants operated a simulated vehicle in six sessions of approximately eight minutes each. The route was identical in all six sessions. It began at a gas station in a small town with pedestrians and multiple traffic lights, then continued onto an 8-shaped, three-lane highway before returning to the starting point. All drives took place in daylight and clear weather conditions. A top-down view of the map, including the route, is shown in \cref{fig:simulator}.
Participants were instructed to follow the route provided by the navigation system, which included both oral directions and a visual display on a dedicated screen. They were asked to drive as they normally would in everyday life, \ie, aiming to reach their destination efficiently while adhering to speed limits and maintaining safe driving behavior to minimize the risk of accidents. It was up to them to decide whether and when to engage the ACC system. 

\subsubsection{Secondary Task} 
In half of the scenarios, one for each level of driving environment complexity, participants had to respond orally to arithmetic calculations dictated by a voice agent. These calculations were triggered based on the vehicle's position along the route. While the specific calculations varied across the three levels of driving environment complexity, their overall difficulty remained comparable. Each calculation involved the addition of two two-digit numbers (\eg, $83+42$).
This secondary task served as a cognitively distracting element, akin to a mentally demanding conversation a driver might have with a passenger or over the phone.

\subsection{Procedure}
Participants were first equipped with electrodes on their left foot to acquire and record electrodermal activity data using a BIOPAC MP160 system (these data were collected as part of another study).
Next, they filled out the pre-test questionnaire. 
The experimenter then introduced the driving simulator. Participants were instructed to drive as naturally as possible while respecting the speed limits of $50$~km/h in the city and $90$~km/h on the highway. They were also informed about the possibility of road construction in certain scenarios, where a $70$~km/h speed limit would apply, and that these would be announced by the navigation system. 

Next, we trained participants on the operation of the driving simulator and on the secondary task through two dedicated training scenarios. The first training one served as a basic tutorial lasting approximately five minutes, conducted on the highway within the CARLA Town 4 map. In this scenario, participants drove alone, with no traffic, and were instructed by the voice agent to perform fundamental tasks such as activating turn signals to change lanes, engaging and disengaging the ACC at specified speeds. The second training scenario followed the same route as in the experiment with low traffic and included a few calculation tasks. This allowed participants to familiarize themselves with both the driving route and the cognitively demanding secondary task before the actual experiment.

Then, participants completed the six test scenarios, each representing different conditions based on a $3$ (driving environment complexity) $\times$ $2$ (cognitive distraction) design. The order of scenarios was counterbalanced across participants. However, two scenarios with the same level of complexity were always presented consecutively, once with cognitive distraction and once without. Half of the participants began with the distraction condition for each level of driving environment complexity, while the other half started without it. 
The three levels of driving environment complexity varied in traffic conditions (\ie, number of vehicles), as shown in \cref{tab:levels_driving_environment_complexity}. The highest level also featured road construction zones along the route, restricting traffic to one or two lanes. However, traffic kept flowing and no standstill sections occurred. Participants were not made aware of the specific conditions for each scenario. After each scenario, participants completed the feedback questionnaire.

\begin{table}[ht]
    \centering
    \caption{Overview of the three levels of driving environment complexity (DEC), detailing the total number of pedestrians, vehicles, and road construction zones present on the route (\cref{fig:simulator}) in each corresponding scenario.}
    \begin{tabular}{l|c|c|c|}
        \hline
        \multicolumn{1}{|l|}{\textbf{DEC} }    & \makecell[c]{\textbf{Nb. of}\\\textbf{pedestrians}}     & \makecell[c]{\textbf{Nb. of}\\\textbf{vehicles}}   & \makecell[c]{\textbf{Nb. of road}\\\textbf{construction}\\\textbf{zones}} \\    
        \hline
        \multicolumn{1}{|l|}{\textbf{\LevelOne} \textbf{(low)}} & $30$ & $37$ & $0$ \\
        \multicolumn{1}{|l|}{\textbf{\LevelTwo} \textbf{(med.)}} & $30$ & $80$ & $0$ \\
        \multicolumn{1}{|l|}{\textbf{\LevelThree} \textbf{(high)}} & $30$ & $80$ & $3$ \\
        \hline
    \end{tabular}
    \label{tab:levels_driving_environment_complexity}
\end{table}

\subsection{Measurements} 
\subsubsection{ACC Use} The use of ACC was evaluated using two measurements: the number of ACC activations and the percentage of time the ACC was engaged while the vehicle was in motion, excluding periods when the vehicle was stationary. Additionally, the feedback questionnaires allowed us to gather the reasons behind participants' decisions to engage, disengage, or refrain from engaging the ACC in each condition. 

\subsubsection{Driving Performance} Driving performance was evaluated based on longitudinal and lateral vehicle control. 
We measured the percentage of time the speed limit was adhered to while the vehicle was in motion, thus excluding periods when the vehicle was stationary. We considered both strict adherence to the speed limit and the effect of allowing a margin of error, defined as a permissible percentage exceeding the limit. For instance, if the speed limit was $90$~km/h and the margin of error was $5\%$, then speeds included in the percentage of time within the speed limit had to remain below $90*(1+5\%)=94.5$~km/h.
Lateral vehicle control was analyzed using the standard deviation of lateral position (SDLP). SDLP was computed after excluding data points corresponding to zero vehicle speed (\eg, when the vehicle was stationary at a traffic light) and periods involving overtaking maneuvers or road changes. These periods were identified by a change in lane or road ID. To account for the full duration of these maneuvers, data were removed from $5$ seconds before to $5$ seconds after each identified lane or road change, as a $10$-second window encompasses $99\%$ of lane changes according to~\citet{Li2021Comprehensive}.
In addition, we also measured the number of lane changes initiated, which serves as a complementary measurement to SDLP for assessing lateral vehicle control. Finally, feedback questionnaires allowed us to gather information on participants' subjective experience of the different driving scenarios and conditions.


\section{Results}
Statistical analyses were conducted using JASP (v0.19.3)~\cite{JASP2025}.
Analyses of variance (ANOVA) with repeated measures were conducted for continuous balanced dependent variables. For continuous unbalanced dependent variables, linear mixed models (LMM) were used. For discrete dependent variables (\eg, count data), generalized linear mixed models (GLMM) with a Poisson distribution were employed.
If significant effects were found, adequate post-hoc comparisons with Bonferroni correction were performed.  In such cases, adjusted p-values are systematically reported in the text.  
All results are reported as statistically significant at $p<.05$. We preferred partial eta squared to partial eta as a measure of effect size for within-subjects design~\cite{Richardson2011Eta}.

\subsection{Insights Provided by the Pre-Test Questionnaire}
\subsubsection{Familiarity with Driving Automation Features} 
All participants reported that they were familiar with the concept of cruise control before taking part in the study. One participant ($3.45\%$) had never driven a vehicle equipped with cruise control, fifteen participants ($51.72\%$) had driven a vehicle equipped with conventional cruise control, and thirteen participants ($44.83\%$) had driven a vehicle equipped with adaptive or intelligent cruise control.
Sixteen participants ($55.17\%$) reported driving a vehicle equipped with cruise control daily, six ($20.69\%$) several times a week, four ($13.79\%$) several times a month, two ($6.90\%$) rarely, and one ($3.45\%$) never.

\subsubsection{Use of Driving Automation Features}
Eighteen participants ($62.07\%$) indicated using driving automation features regularly, in particular seven ($24.14\%$) responding `completely' and eleven ($37.93\%$) `rather yes' on a five-point scale from `Not at all' to `Completely' to the question `Do you regularly activate driving assistance features (\eg, cruise control, lane-keeping assist)?'. The main reasons given were comfort and convenience, maintaining a constant speed or adhering to speed limits, and managing fuel consumption. 
Nine participants ($31.03\%$) indicated not using driving automation features regularly, in particular four ($13.79\%$) responding `not at all' and five ($17.24\%$) `rather no'. Their main reasons were a dislike of driving automation features, a preference for being in full control of the vehicle, a belief that they did not need them, or simply forgetting to engage them. 
Two participants ($6.90\%$) gave a neutral response, mainly because they only use these features on highways. 

\subsubsection{Attitudes toward Driving Automation Features}
Twenty-six participants ($89.66\%$) stated that they were in favor of driving automation features, with fourteen ($48.28\%$) responding `completely' and twelve ($41.38\%$) `rather yes'. The main reasons mentioned were comfort and convenience, increased safety and reduced risk of accidents, reduced fatigue, and the ability to focus more on other tasks.
Three participants ($10.34\%$) responded `neither yes nor no' stating that they prefer to remain in full control of the vehicle or do not use the features anyway.

Finally, twenty-two participants ($75.86\%$) stated that they favored (semi-)autonomous vehicles (namely, vehicles at SAE Levels 3, 4 and 5), with seven ($24.14\%$) responding `completely' and fifteen ($51.72\%$) `rather yes'. Similarly to driving automation features, the main reasons included comfort and convenience, increased safety and reduced risk of accidents. Additionally, participants mentioned time management—allowing them to engage in other activities while driving—and a smoother traffic flow.
Four participants ($13.79\%$) were not in favor of (semi-)autonomous vehicles, with one ($3.45\%$) responding `not at all' and three ($10.34\%$) `rather no'. The main reasons cited were the desire to maintain full control of the vehicle, a love of driving, and the importance of retaining driving capacity.
Three participants ($10.34\%$) gave a neutral response, stating that their opinion depended on the intended use of the vehicle or that they did not fully trust autonomous driving.

\subsection{Effects of Driving Environment Complexity and Cognitive Distraction on ACC Use (RQ1 \& RQ2)}
\subsubsection{Number of ACC Activations}
A generalized linear mixed model with a Poisson distribution and log link function was fitted to analyze the number of ACC activations, based on $174$ observations from $29$ participants each completing the $6$ scenarios ($29 \times 6 = 174$). It included fixed effects of driving environment complexity and cognitive distraction, with participant as a random intercept to account for individual differences. 
Results showed no significant main effect of driving environment complexity ($\chi^2(2)=4.348, p=.114$) or cognitive distraction ($\chi^2(1)=.442, p=.506$) on ACC activations. Additionally, there was no significant interaction between environment complexity and distraction ($\chi^2(2)=.544, p=.762$).

\begin{table*}[ht]
    \centering
    \caption{Average number of ACC activations and percentage of ACC engagement time (SD; Min-Max) for the $N=29$ participants in each of the six scenarios, \ie, in the different driving environment complexities (DEC), with and without cognitive distraction.}
    \begin{tabular}{l|c|c|c|c|}
        \cline{2-5}
                                                & \multicolumn{2}{c|}{\textbf{Number of ACC activations}}    & \multicolumn{2}{c|}{\textbf{Percentage of ACC engagement time}} \\
        \hline
        \multicolumn{1}{|l|}{\textbf{DEC} }     & \textbf{No distraction}       & \textbf{Distraction}       & \textbf{No distraction}       & \textbf{Distraction} \\    
        \hline
         \multicolumn{1}{|l|}{\textbf{\LevelOne}} & $3.4$ ($1.9$; $1$-$9$)     & $3.1$ ($1.4$; $1$-$9$)  & $51.6$ ($18.2$; $0.4$-$80.3$) & $53.4$ ($15.4$; $7.9$-$83.0$) \\
         \multicolumn{1}{|l|}{\textbf{\LevelTwo}} & $3.2$ ($2.2$; $0$-$11$)    & $3.3$ ($1.1$; $1$-$5$)  & $50.3$ ($21.9$; $0.0$-$83.1$) & $49.8$ ($15.0$; $19.3$-$82.3$)\\
         \multicolumn{1}{|l|}{\textbf{\LevelThree}} & $4.0$ ($1.9$; $0$-$11$)    & $3.7$ ($2.2$; $0$-$10$) & $43.3$ ($19.2$; $0.0$-$69.1$) & $41.7$ ($24.1$; $0.0$-$72.9$) \\
        \hline
    \end{tabular}
    \label{tab:ACC_use}
\end{table*}

\subsubsection{Percentage of ACC Engagement Time}
A two-way repeated-measures ANOVA was conducted to examine the effects of driving environment complexity (\LevelOne, \LevelTwo, \LevelThree) and cognitive distraction (present or absent) on the percentage of ACC engagement time. Mauchly's test of sphericity was significant for both driving environment complexity ($\chi^2(2)=6.573, p=.037$) and the interaction of the two factors ($\chi^2(2)=10.654, p=.005$), indicating that the assumption of sphericity was violated. A Huynh-Feldt correction was applied, with epsilon values of $\epsilon=.867$ for driving environment complexity and $\epsilon=.788$ for the interaction. The standard p-value without correction was used for cognitive distraction, as the assumption of sphericity applies only to factors with three or more levels.

The analysis revealed a significant main effect of driving environment complexity ($F(1.734,48.546)=4.391, p=.022, \eta_p^2=.136$), with a medium effect size (see \cref{tab:ACC_use}). Post-hoc comparisons using the Bonferroni correction revealed that the percentages of ACC engagement time at \LevelThree of driving environment complexity was significantly smaller than at \LevelOne ($p=.022$), while no significant difference was found between \LevelTwo and the two other levels ($p=1.0$ for \LevelOne and $p=.248$ for \LevelThree).  
The main effect of cognitive distraction, as well as the interaction between driving environment complexity and cognitive distraction, were not significant ($F(1,28)=.005, p=.947, \eta_p^2=.0001$ and $F(1.576,44.115)=.271, p=.711, \eta_p^2=.010$, respectively). In other words, ACC engagement time is only influenced significantly by the level of driving environment complexity but not by cognitive distraction.

\subsubsection{Rationale for ACC Use based on Feedback Questionnaires}
Participants stated that they activated ACC in all scenarios whenever traffic conditions allowed and to assist with maintaining speed limits. They also reported using ACC during cognitively demanding tasks to help manage all tasks. Additionally, participants indicated activating ACC while driving on the highway.
In contrast, participants reported not activating ACC in any scenario when traffic conditions were unsuitable. In particular, they found it challenging to activate ACC in complex driving environments, especially when cognitively distracted. Additionally, they refrained from using ACC in urban areas or situations requiring frequent braking, such as approaching traffic lights.
Finally, participants indicated that they deactivated ACC in all scenarios when traffic conditions required it. They also disengaged ACC at highway exits to brake.

\subsection{Impact of ACC Use on Driving Performance (RQ3 \& RQ4)}
\subsubsection{Percentage of Time within the Speed Limit}
Each of the $29$ participants completed the $6$ scenarios ($3$ driving environment complexity $\times$ $2$ cognitive distraction). Within each scenario, they were free to activate and deactivate ACC as often as they wished. Thus, we expected to process $29 \times 6 \times 2 = 348$ observations. However, $6$ observations were missing because participants had not engaged the ACC at all during an entire scenario. Such missing values led to unbalanced data when analyzing the effect of ACC use on the percentage of time participants were driving under the speed limit. 
Therefore, a linear mixed model was fitted using restricted maximum likelihood (REML), with Satterthwaite's method, with $342$ observations from $29$ participants.  
The model included driving environment complexity, cognitive distraction, and ACC as fixed effects, with a random intercept for participants to account for individual variability.

\begin{figure*}
    \centering
    \includegraphics[width=\linewidth]{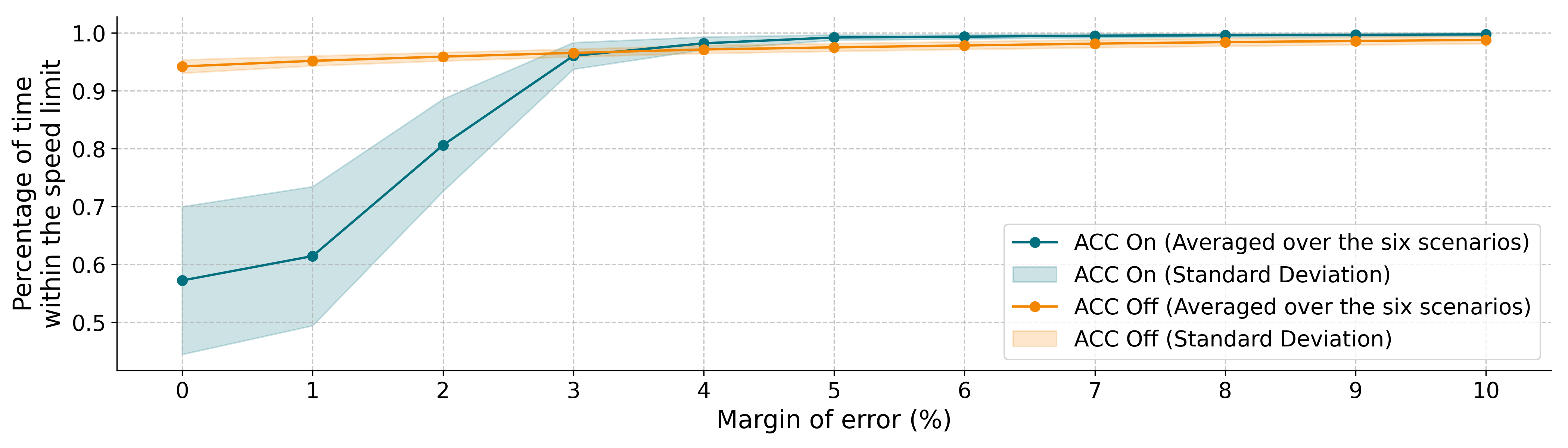}
    \Description{This graph shows the percentage of time within the speed limit for margins of error ranging from zero percent to ten percent.}
    \caption{Percentage of time within the speed limit, averaged across the six scenarios (three levels of driving environment complexity $\times$ two levels of cognitive distraction), as a function of the allowed margin of error on the speed limit, when the ACC is engaged (ACC On, in blue) and disengaged (ACC Off, in orange).
    When the margin of error is $0\%$, the percentage of time within the speed limit is lower when the ACC is engaged compared to when it is not. However, as the margin of error increases, this percentage rises more notably when the ACC is engaged, eventually surpassing the percentage observed when the ACC is disengaged. 
    This effect can be attributed to the ACC's operation, which aims to maintain the set speed but may occasionally exceed it slightly, or to instances where the ACC set speed is slightly above the speed limit.}
    \label{fig:proportion_within_limit}
\end{figure*}

A margin of error was considered when comparing the vehicle's speed to the speed limit. In the driving simulator, when the ACC was set to a given speed, the system attempted to regulate the vehicle's speed accordingly, but some variations leading to higher speeds still occurred. As a result, when no margin of error was allowed, the percentage of time within the speed limit while ACC was engaged appeared unexpectedly low, as shown in~\cref{fig:proportion_within_limit}. However, when a $5\%$ margin of error was permitted, the percentage of time within the speed limit was higher when ACC was engaged than when it was disengaged, meaning that vehicles with ACC engaged were more often within the speed limit than vehicles with ACC disengaged.

We fitted linear mixed models with margins of error of $0\%$ and $5\%$. 
When no margin of error was considered (see \cref{tab:proportion_within_limit_no_margin}), the results showed significant effects of driving environment complexity ($F(2,308.61)=30.308, p<.001$), cognitive distraction ($F(1,308.58)=11.501, p<.001$), and ACC use ($F(1,308.67)=583.321, p<.001$) on the percentage of time within the speed limit.
The percentage of time within the speed limit increased as driving environment complexity increased. Post-hoc pairwise comparisons (using contrasts) revealed significant differences between all levels ($\LevelOne-\LevelTwo$: $b=-.064, SE=.019, z=-3.396, p=.002$; $\LevelTwo-\LevelThree$: $b=-.082, SE=.019, z=-4.344, p<.001$; $\LevelOne-\LevelThree$: $b=-.146, SE=.019, z=-7.769, p<.001$).
Moreover, participants drove more time within the speed limit when cognitively distracted compared to when they were not distracted, and when ACC was disengaged compared to when it was engaged.

\aptLtoX{\begin{table*}[ht]
    \centering
    \caption{Average percentage of time within the speed limit (SD; Min-Max) with $0\%$ and $5\%$ margins of error for the $N=29$ participants when the ACC was engaged or disengaged across the six scenarios (three levels of driving environment complexity (DEC) $\times$ two levels of cognitive distraction).
    In other words, this corresponds to the conditional probability that speed is within the limit with a margin of error, given whether ACC is on or off, whether the driver is distracted or not, and whether driving environment complexity is at \LevelOne, \LevelTwo, or \LevelThree (\eg, $P(\text{speed < (1+margin of error)$*$speed limit} | \text{ACC=off, distraction=without, DEC=\LevelOne})$ for the top left cell of the table).}
    \label{tab:proportion_within_limit}
    \label{tab:proportion_within_limit_margin_5}
    \label{tab:proportion_within_limit_no_margin}
   
    \begin{tabular}{l|c|c|c|c|}
     \multicolumn{5}{c|}{\textbf{(a) No margin of error}}\\
 
        \cline{2-5}
                                                & \multicolumn{4}{c|}{\textbf{Percentage of time within the speed limit with no margin of error}} \\
        \cline{2-5}                             
                                                & \multicolumn{2}{c|}{\textbf{No distraction}}                  & \multicolumn{2}{c|}{\textbf{Distraction}}\\
        \hline
        \multicolumn{1}{|l|}{\textbf{DEC} }     & \textbf{ACC Off}              & \textbf{ACC On}               & \textbf{ACC Off}              & \textbf{ACC On} \\    
        \hline
        \multicolumn{1}{|l|}{\textbf{\LevelOne}} & $92.1$ ($14.8$; $34.9$-$100$)   & $39.4$ ($19.6$; $0.0$-$100$)   & $95.0$ ($9.5$; $53.9$-$100$)   & $47.7$ ($23.2$; $6.6$-$100$) \\
        \multicolumn{1}{|l|}{\textbf{\LevelTwo}} & $93.3$ ($12.1$; $47.2$-$100$)   & $50.1$ ($15.3$; $27.3$-$81.8$) & $95.0$ ($8.8$; $60.4$-$100$)   & $61.1$ ($26.3$; $28.2$-$100$) \\
        \multicolumn{1}{|l|}{\textbf{\LevelThree}} & $94.6$ ($10.0$; $52.3$-$100$)   & $67.8$ ($18.6$; $33.4$-$100$)  & $95.3$ ($9.6$; $55.3$-$100$)   & $77.1$ ($18.9$; $31.5$-$100$) \\
        \hline
    \end{tabular}

    \begin{tabular}{l|c|c|c|c|}
        \multicolumn{5}{c|}{\textbf{(b) $5\%$ margin of error}}\\
        \cline{2-5}
                                                & \multicolumn{4}{c|}{\textbf{Percentage of time within the speed limit with a $5\%$ margin of error}} \\
        \cline{2-5}
                                                & \multicolumn{2}{c|}{\textbf{No distraction}}                  & \multicolumn{2}{c|}{\textbf{Distraction}}\\
        \hline
        \multicolumn{1}{|l|}{\textbf{DEC} }     & \textbf{ACC Off}              & \textbf{ACC On}               & \textbf{ACC Off}              & \textbf{ACC On} \\    
        \hline
        \multicolumn{1}{|l|}{\textbf{\LevelOne}} & $96.3$ ($12.5$; $41.3$-$100$)   & $98.3$ ($7.4$; $60.3$-$100$)   & $97.5$ ($7.5$; $62.3$-$100$)   & $99.3$ ($2.3$; $89.0$-$100$) \\
        \multicolumn{1}{|l|}{\textbf{\LevelTwo}} & $97.0$ ($8.6$; $57.7$-$100$)    & $99.3$ ($2.3$; $88.6$-$100$)   & $97.7$ ($7.0$; $64.5$-$100$)   & $99.1$ ($3.2$; $83.3$-$100$) \\
        \multicolumn{1}{|l|}{\textbf{\LevelThree}} & $98.3$ ($6.6$; $64.8$-$100$)    & $99.6$ ($1.2$; $94.4$-$100$)   & $98.1$ ($6.6$; $65.8$-$100$)   & $99.7$ ($1.1$; $95.0$-$100$) \\
        \hline
    \end{tabular}
\end{table*}}{\begin{table*}[ht]
    \centering
    \caption{Average percentage of time within the speed limit (SD; Min-Max) with $0\%$ and $5\%$ margins of error for the $N=29$ participants when the ACC was engaged or disengaged across the six scenarios (three levels of driving environment complexity (DEC) $\times$ two levels of cognitive distraction).
    In other words, this corresponds to the conditional probability that speed is within the limit with a margin of error, given whether ACC is on or off, whether the driver is distracted or not, and whether driving environment complexity is at \LevelOne, \LevelTwo, or \LevelThree (\eg, $P(\text{speed < (1+margin of error)$*$speed limit} | \text{ACC=off, distraction=without, DEC=\LevelOne})$ for the top left cell of the table).}
    \label{tab:proportion_within_limit}
    
    \begin{subtable}{\textwidth}
    \centering
    \caption{No margin of error}
    \label{tab:proportion_within_limit_no_margin}
    \begin{tabular}{l|c|c|c|c|}
        \cline{2-5}
                                                & \multicolumn{4}{c|}{\textbf{Percentage of time within the speed limit with no margin of error}} \\
        \cline{2-5}                             
                                                & \multicolumn{2}{c|}{\textbf{No distraction}}                  & \multicolumn{2}{c|}{\textbf{Distraction}}\\
        \hline
        \multicolumn{1}{|l|}{\textbf{DEC} }     & \textbf{ACC Off}              & \textbf{ACC On}               & \textbf{ACC Off}              & \textbf{ACC On} \\    
        \hline
        \multicolumn{1}{|l|}{\textbf{\LevelOne}} & $92.1$ ($14.8$; $34.9$-$100$)   & $39.4$ ($19.6$; $0.0$-$100$)   & $95.0$ ($9.5$; $53.9$-$100$)   & $47.7$ ($23.2$; $6.6$-$100$) \\
        \multicolumn{1}{|l|}{\textbf{\LevelTwo}} & $93.3$ ($12.1$; $47.2$-$100$)   & $50.1$ ($15.3$; $27.3$-$81.8$) & $95.0$ ($8.8$; $60.4$-$100$)   & $61.1$ ($26.3$; $28.2$-$100$) \\
        \multicolumn{1}{|l|}{\textbf{\LevelThree}} & $94.6$ ($10.0$; $52.3$-$100$)   & $67.8$ ($18.6$; $33.4$-$100$)  & $95.3$ ($9.6$; $55.3$-$100$)   & $77.1$ ($18.9$; $31.5$-$100$) \\
        \hline
    \end{tabular}
    \end{subtable}

    \bigskip

    \begin{subtable}{\textwidth}
    \centering
    \caption{$5\%$ margin of error}
    \label{tab:proportion_within_limit_margin_5}
    \begin{tabular}{l|c|c|c|c|}
        \cline{2-5}
                                                & \multicolumn{4}{c|}{\textbf{Percentage of time within the speed limit with a $5\%$ margin of error}} \\
        \cline{2-5}
                                                & \multicolumn{2}{c|}{\textbf{No distraction}}                  & \multicolumn{2}{c|}{\textbf{Distraction}}\\
        \hline
        \multicolumn{1}{|l|}{\textbf{DEC} }     & \textbf{ACC Off}              & \textbf{ACC On}               & \textbf{ACC Off}              & \textbf{ACC On} \\    
        \hline
        \multicolumn{1}{|l|}{\textbf{\LevelOne}} & $96.3$ ($12.5$; $41.3$-$100$)   & $98.3$ ($7.4$; $60.3$-$100$)   & $97.5$ ($7.5$; $62.3$-$100$)   & $99.3$ ($2.3$; $89.0$-$100$) \\
        \multicolumn{1}{|l|}{\textbf{\LevelTwo}} & $97.0$ ($8.6$; $57.7$-$100$)    & $99.3$ ($2.3$; $88.6$-$100$)   & $97.7$ ($7.0$; $64.5$-$100$)   & $99.1$ ($3.2$; $83.3$-$100$) \\
        \multicolumn{1}{|l|}{\textbf{\LevelThree}} & $98.3$ ($6.6$; $64.8$-$100$)    & $99.6$ ($1.2$; $94.4$-$100$)   & $98.1$ ($6.6$; $65.8$-$100$)   & $99.7$ ($1.1$; $95.0$-$100$) \\
        \hline
    \end{tabular}
    \end{subtable}
\end{table*}}

When a $5\%$ margin of error was considered (see \cref{tab:proportion_within_limit_margin_5}), the results showed a significant main effect of ACC use ($F(1,308.67)=10.188, p=.002$) on the percentage of time within the speed limit. However, there was no significant main effect of driving environment complexity ($F(2,308.66)=1.282, p=.279$) or cognitive distraction ($F(1,308.65)=.778, p=.379$), and no significant interaction. 
As observed in~\cref{fig:proportion_within_limit} and \cref{tab:proportion_within_limit_margin_5}, when accounting for a $5\%$ margin of error, the percentage of time within the speed limit was higher when ACC was engaged compared to when it was disengaged. Although the difference was not significant, the percentage of time within the speed limit tended to increase as driving environment complexity increased and when participants were cognitively distracted, compared to when they were not distracted.

\subsubsection{Standard deviation of lateral position (SDLP)}
A linear mixed model was fitted using restricted maximum likelihood (REML), with Satterthwaite's method, based on the same $342$ observations. The model included driving environment complexity, cognitive distraction, and ACC as fixed effects, with a random intercept for participants to account for individual variability. Results showed significant effects of driving environment complexity ($F(2,301.87)=5.966, p=.003$), cognitive distraction ($F(1,301.82)=8.002, p=.005$), and ACC use ($F(1,302.01)=48.776, p<.001$) on SDLP. A significant two-way interaction between driving environment complexity and ACC use was found ($F(2,301.87)=3.154, p=.044$). No other significant interaction was observed. 

\Cref{tab:SDLP} indicates that SDLP was lower when participants were cognitively distracted than when they were not, and that it was also lower when ACC was engaged than when it was disengaged. 
Post-hoc pairwise comparisons (using contrasts) were conducted to explore the effect of driving environment complexity on SDLP, averaging over cognitive distraction and ACC use.  
Results showed that SDLP was significantly lower at $\LevelOne$ than at $\LevelTwo$ ($b=-.090, SE=.026, z=-3.444, p=.002$) of driving environment complexity. 
No significant difference was however observed between $\LevelThree$ and the two other levels ($\LevelOne$: $b=-.050, SE=.026, z=-1.934, p=.159$; $\LevelTwo$: $b=.039, SE=.026, z=1.500, p=.401$).
We also analyzed the significant interaction between the driving environment complexity and ACC use, comparing ACC on and ACC off within each level of environment complexity. Results showed that ACC significantly reduced SDLP in $\LevelOne$ ($b=.147, SE=.037, Z=4.001, p<.001$) and $\LevelTwo$ of driving environment complexity ($b=.216, SE=.037, z=5.821, p<.001$). In $\LevelThree$, the effect was smaller and did not reach significance ($b=.085, SE=.037, z=2.275, p=.069$).

\begin{table*}[ht]
    \centering
    \caption{Average standard deviation of lateral position (SDLP) (SD; Min-Max) for the $N=29$ participants when the ACC was engaged or disengaged across the six scenarios, \ie, in the three levels of driving environment complexity (DEC), with and without cognitive distraction.}
    \begin{tabular}{l|c|c|c|c|c|c|c|c|}
        \cline{2-5}
                                                        & \multicolumn{4}{c|}{\textbf{SDLP (m)}} \\
        \cline{2-5}
                                                        & \multicolumn{2}{c|}{\textbf{No distraction}}                & \multicolumn{2}{c|}{\textbf{Distraction}} \\
        \hline
        \multicolumn{1}{|l|}{\textbf{DEC} }             & \textbf{ACC Off}            & \textbf{ACC On}               & \textbf{ACC Off}             & \textbf{ACC On} \\    
        \hline
        \multicolumn{1}{|l|}{\textbf{\LevelOne}}        & $.67$ ($.21$; $.38$-$1.2$) & $.51$ ($.12$; $.04$-$.68$) & $.64$ ($.18$; $.38$-$1.1$) & $.51$ ($.08$; $.37$-$.71$) \\
        \multicolumn{1}{|l|}{\textbf{\LevelTwo}}        & $.86$ ($.46$; $.35$-$2.4$) & $.61$ ($.23$; $.34$-$1.5$) & $.70$ ($.21$; $.31$-$1.3$) & $.52$ ($.08$; $.32$-$.68$) \\
        \multicolumn{1}{|l|}{\textbf{\LevelThree}}      & $.69$ ($.24$; $.39$-$1.6$) & $.60$ ($.13$; $.36$-$1.0$) & $.65$ ($.22$; $.33$-$1.4$) & $.57$ ($.14$; $.34$-$1.0$) \\
        \hline
    \end{tabular}
    \label{tab:SDLP}
\end{table*}

\subsubsection{Number of Lane Changes}
A generalized linear mixed model with a Poisson distribution and log link function was fitted, based on the same $342$ observations. The results revealed a significant main effect of driving environment complexity ($\chi^2(2)=57.354, p<.001$), but no significant effect of cognitive distraction ($\chi^2(1)=1.310, p=.252$) or ACC use ($\chi^2(1)=2.111, p=.146$) on the number of lane changes. A significant three-way interaction between driving environment complexity, cognitive distraction, and ACC use was found ($\chi^2(2)=11.648, p=.003$). However, no significant two-way interactions were observed.
We observe in~\cref{tab:lane_changes} that the number of lane changes increased with greater driving environment complexity, likely due to higher traffic levels and road constructions requiring more overtaking maneuvers. Post-hoc pairwise comparisons (using contrasts) revealed significant difference between $\LevelOne$ and the two others levels ($\LevelTwo$: $b=-1.700, SE=.350, z=-4.860, p<.001$; $\LevelThree$: $b=-2.590, SE=.375, z=-6.905, p<.001$), but no significant difference between $\LevelTwo$ and $\LevelThree$ ($b=-.890, SE=.374, z=-2.378, p=.052$).
We also analyzed the significant three-way interaction between driving environment complexity, cognitive distraction, and ACC use. Results showed that participants significantly changed lane more often when the ACC was activated than when it was deactivated at $\LevelTwo$ of driving environment complexity when they were not distracted. In all other conditions, ACC use had no significant effect on the number of lane changes.

\begin{table*}[ht]
    \centering
    \caption{Average number of lane changes (SD; Min-Max) for the $N=29$ participants when the ACC was engaged or disengaged across the six scenarios, \ie, in the three levels of driving environment complexity (DEC), with and without cognitive distraction.}
    \begin{tabular}{l|c|c|c|c|}
        \cline{2-5}
                                                & \multicolumn{4}{c|}{\textbf{Number of lane changes}} \\
        \cline{2-5}
                                                & \multicolumn{2}{c|}{\textbf{No distraction}}                & \multicolumn{2}{c|}{\textbf{Distraction}}\\
        \hline
        \multicolumn{1}{|l|}{\textbf{DEC} }     & \textbf{ACC Off}            & \textbf{ACC On}               & \textbf{ACC Off}             & \textbf{ACC On} \\    
        \hline
        \multicolumn{1}{|l|}{\textbf{\LevelOne}} & $6.1$ ($2.9$; $2$-$13$)   & $7.0$ ($5.3$; $0$-$20$)    & $5.3$ ($3.7$; $1$-$14$)   & $6.3$ ($4.2$; $0$-$15$) \\
        \multicolumn{1}{|l|}{\textbf{\LevelTwo}} & $6.7$ ($4.3$; $1$-$17$)   & $9.4$ ($5.4$; $2$-$20$)    & $8.6$ ($5.9$; $0$-$25$)   & $7.2$ ($4.2$; $1$-$19$) \\
        \multicolumn{1}{|l|}{\textbf{\LevelThree}} & $9.3$ ($4.9$; $1$-$24$)   & $8.3$ ($5.2$; $0$-$18$)    & $9.0$ ($4.7$; $1$-$18$)   & $8.6$ ($6.6$; $0$-$22$) \\
        \hline
    \end{tabular}
    \label{tab:lane_changes}
\end{table*}

\subsubsection{Perceived Safety based on Feedback Questionnaires}
Regarding the subjective ratings to the statement `It was complicated to drive safely' (on a seven-point scale from (1) `strongly disagree' to (7) `strongly agree'), a two-way repeated-measures ANOVA revealed a significant main effect of driving environment complexity ($F(2,56)=7.974, p<.001, \eta_p^2=.222$), with a large effect size. Post-hoc comparisons with Bonferroni correction showed that \LevelTwo and \LevelThree of driving environment complexity were rated as being more complicated to drive safely than \LevelOne (respectively, $p=.042$ and $p=.002$), but there was no significant difference between \LevelTwo and \LevelThree. There was also a large main effect of cognitive distraction ($F(1,28)=18.06, p<.001, \eta_p^2=.392$) with people finding it more complicated to drive safely when cognitively distracted compared to when they were not. Finally, there was no significant interaction between the two factors (see \cref{tab:complicated}).

\begin{table}[ht]
    \centering
    \caption{Average subjective ratings (SD; Min-Max) to the statement `It was complicated to drive safely' (on a seven-point scale from (1) `strongly disagree' to (7) `strongly agree') for the $N=29$ participants in each of the six scenarios, \ie, in the different driving environment complexities (DEC), with and without cognitive distraction.}
    \begin{tabular}{l|c|c|}
        \cline{2-3}
                                                    & \multicolumn{2}{c|}{\textbf{`It was complicated to drive safely'}} \\
        \hline
        \multicolumn{1}{|l|}{\textbf{DEC} }         & \textbf{No distraction}       & \textbf{Distraction} \\    
        \hline
         \multicolumn{1}{|l|}{\textbf{\LevelOne}}   & $2.6$ ($1.8$; $1$-$7$)         & $3.4$ ($1.6$; $1$-$6$) \\
         \multicolumn{1}{|l|}{\textbf{\LevelTwo}}   & $3.3$ ($1.9$; $1$-$7$)         & $3.9$ ($1.8$; $1$-$7$) \\
         \multicolumn{1}{|l|}{\textbf{\LevelThree}} & $3.4$ ($1.7$; $1$-$7$)         & $4.9$ ($1.7$; $1$-$7$) \\
        \hline
    \end{tabular}
    \label{tab:complicated}
\end{table}


\section{Discussion}
Our results indicate that driving environment complexity influences the percentage of ACC engagement time, but not the number of ACC activations (\textbf{RQ1}).
Regardless of cognitive distraction conditions, the percentage of time ACC is engaged was lower in the most complex driving environment compared to the less complex one.
In line with survey studies~\cite{Wu2015Drivers}, participants reported that they preferentially activated ACC on highways but refrained from using it in urban areas where frequent braking was required, notably at traffic lights.
Overall, it seems that cognitive distraction does not influence the absolute ACC use, either in terms of the number of activations or overall engagement time, even if people may feel that the more demanding situations (\ie, most complex environment coupled with a secondary task) make the activation more challenging (\textbf{RQ2}).

Additionally, our study suggests that ACC use influences driving performance (\textbf{RQ3} and \textbf{RQ4}). While ACC use does not appear to affect the number of lane changes, it impacts both the percentage of time the speed limit is adhered to and the SDLP. 
Specifically, regarding longitudinal control, whether ACC has a positive or negative effect on speed limit compliance depends on the margin of error allowed. Indeed, if no margin of error is permitted, and we consider only the percentage of time the vehicle's speed is strictly less than or equal to the speed limit, then driving with ACC engaged leads to lower speed limit compliance than driving without it. This effect may be explained by the ACC's implementation, which aims to maintain a set speed but may occasionally exceed it slightly, or by situations where the set speed itself is slightly above the speed limit.
In contrast, if a $5\%$ margin of error is allowed (meaning that the vehicle's speed can slightly exceed the limit), then driving with ACC improves speed limit compliance.  
Therefore, allowing a margin of error cancels this effect, resulting in nearly total compliance within the speed limit plus the margin of error. Notably, a $5\%$ tolerance is commonly accepted by speed cameras in real-world settings.
Note that participants reported in feedback questionnaires that they activated ACC to help them maintain speed limits. This was also one of the main reasons cited in the pre-test questionnaire for using driving automation features regularly in real life. 
In the literature, the effect of ACC on speed remains inconclusive. For instance, \citet{Hoedemaeker1998Behavioural} found that drivers in a simulator study were driving faster with ACC activated, while \citet{Vollrath2011TheInfluence} observed lower maximum speed and fewer speed limit violations when driving with ACC compared to manual driving. Meanwhile, \citet{BianchiPiccinini2014Drivers} found no statistically significant effect of ACC on traveling speed. However, as shown in our study, results highly depend on the way ACC is implemented and on the measures used.
Besides, when considering strict compliance (no margin of error), speed limit adherence increases with driving environment complexity and cognitive distraction, regardless of ACC use. With a $5\%$ margin of error, no significant differences are observed.
Regarding lateral control, SDLP was lower when ACC was engaged compared to when it was disengaged, suggesting that ACC use improves lateral driving performance. SDLP was also lower when participants were cognitively distracted than when they were not, a finding consistent with previous studies~\cite{Liang2010Combining,Ranney2008Driver}. Additionally, lateral control was better (\ie, lower SDLP) in the lowest driving environment complexity compared to the medium complexity level. Notably, a significant interaction was found between driving environment complexity and ACC use, suggesting that lateral control was improved by ACC use in the two lowest levels of environment complexity, but not in the most complex one.

As for lane changes, the complexity of the driving environment had a significant effect, with more lane changes occurring in environments with higher traffic density, likely due to more frequent overtaking maneuvers. 
Neither cognitive distraction nor ACC use had a significant effect on the number of lane changes. However, a significant three-way interaction was observed between driving environment complexity, cognitive distraction, and ACC use. A higher number of lane changes was observed when ACC was engaged compared to when it was disengaged, but only in the medium-complexity environment and in the absence of distraction. No significant differences were found in the other conditions.

Our findings underscore the importance of adopting a \emph{dynamic operational level of automation} starting as early as SAE Level 1, as even a feature like ACC was activated and deactivated multiple times during an eight-minute route. Moreover, the study supports the idea that adaptive automation, dynamically adjusting the level of automation, should account for both driving environment complexity and driver state.
While cognitive distraction, unlike driving environment complexity, did not significantly influence ACC use in this study, participants reported that both factors affected their perception of driving safety.
Although participants did not significantly modify their ACC use in response to cognitive distraction, automatically adapting the automation level based on driver state might still improve driving performance.


\section{Limitations}
We acknowledge several limitations to our study.
First, we operationalized driver state, driving environment complexity, and driving automation features by focusing specifically on cognitive distraction, traffic density, and ACC, respectively. While this constitutes an important first step, future research should explore other driver states (\eg, drowsiness or visual distraction), additional aspects of driving environment complexity (\eg, weather conditions or road types), and other driving automation features (\eg, lane-keeping assistance (LKA)) to assess whether similar conclusions can be drawn.
Furthermore, traffic density was constrained by the computational limits of the system running CARLA, as higher densities increased the risk of simulator crashes. 
Besides, it is worth noting that the ACC system used here could only be disengaged using a designated button, as pressing the brake pedal neither deactivated ACC nor initiated braking while ACC was engaged. This differs from real-world ACC implementations, where braking typically deactivates the system.
Second, participants were not selected based on age, background, or driving experience. As these factors may also influence automation use and driving behavior, their impact should be considered in future studies to support the development of adaptive automation that is commensurate with experience or other drivers' characteristics.
Of course, as with all simulator studies, the question of transferability, reliability, and validity of the results to real-world driving conditions arises. Although driving simulators allow for controlled experimental conditions, conclusions drawn in a simulated environment may differ from the ones drawn in the real world.


\section{Conclusion}
With our simulator study, we investigated whether and how driving environment complexity and driver state influence reliance on driving automation features (\textbf{RQ1} \& \textbf{RQ2}), and how using such features may impact driving performance (\textbf{RQ3} \& \textbf{RQ4}).
$N = 29$ participants drove in six different conditions, consisting of three levels of driving environment complexity and two levels of cognitive distraction, with the ability to activate or deactivate ACC at any time.  

We found that (\textbf{RQ1}) the overall time of ACC engagement was lower in complex driving environments than in less complex ones, while the number of ACC activations itself was not significantly affected by the environment. Additionally, (\textbf{RQ2}) we did not observe any significant effect of cognitive distraction on ACC use. 
Regarding driving performance, our study revealed that ACC use had an effect on the number of lane changes only in the medium-complexity environment and in the absence of cognitive distraction. ACC use led to lower SDLP and thus improved lateral control in the two lowest levels of environment complexity, no matter if participants were distracted or not. Furthermore, ACC use did have a negative or positive impact on compliance with speed limits, depending on whether a strict speed limit or a margin of error was considered.
Additionally, we found that (\textbf{RQ3}) more lane changes occurred in more complex driving environments. The lateral control (\ie, SDLP) was better in the least complex driving environment. 
For strict speed limit compliance, adherence was higher without ACC and increased with driving environment complexity, regardless of ACC use. However, with a $5\%$ margin of error, speed limit adherence was higher with ACC than without, with no significant effect of driving environment complexity.
Finally, (\textbf{RQ4}) cognitive distraction did not affect the number of lane changes. However, lateral control was better with ACC and with cognitive distraction than in other situations, while strict speed limit compliance was better without ACC and with cognitive distraction. With a $5\%$ margin of error, speed limit adherence was higher with ACC but with no significant effect of cognitive distraction.

In conclusion, the present study suggests that to develop adaptive automation features that vehicle users rely on, it is essential to consider both the driving environment and the driver state. This will help ensure that driving automation systems can dynamically adjust to varying conditions and driver needs to enhance both safety and comfort.


\begin{acks}
The work by A. Halin was supported by the SPW EER, Wallonia, Belgium under grant n°2010235 (ARIAC by \href{https://www.digitalwallonia.be/en/}{DIGITALWALLONIA4.AI}). 
The authors thank AISIN Europe for providing access to their driving simulator, and specifically acknowledge Richard Virlouvet and Frédéric Burguet for their assistance and support with the simulator.
\end{acks}

\bibliographystyle{ACM-Reference-Format}
\bibliography{main}


\begin{thebibliography}{50}


\ifx \showCODEN    \undefined \def \showCODEN     #1{\unskip}     \fi
\ifx \showISBNx    \undefined \def \showISBNx     #1{\unskip}     \fi
\ifx \showISBNxiii \undefined \def \showISBNxiii  #1{\unskip}     \fi
\ifx \showISSN     \undefined \def \showISSN      #1{\unskip}     \fi
\ifx \showLCCN     \undefined \def \showLCCN      #1{\unskip}     \fi
\ifx \shownote     \undefined \def \shownote      #1{#1}          \fi
\ifx \showarticletitle \undefined \def \showarticletitle #1{#1}   \fi
\ifx \showURL      \undefined \def \showURL       {\relax}        \fi
\providecommand\bibfield[2]{#2}
\providecommand\bibinfo[2]{#2}
\providecommand\natexlab[1]{#1}
\providecommand\showeprint[2][]{arXiv:#2}

\bibitem[Ahlstr{\"o}m et~al\mbox{.}(2018a)]%
        {Ahlstrom2018TheEffect}
\bibfield{author}{\bibinfo{person}{Christer Ahlstr{\"o}m},
  \bibinfo{person}{Anna Anund}, \bibinfo{person}{Carina Fors}, {and}
  \bibinfo{person}{Torbj{\"o}rn {\AA}kerstedt}.}
  \bibinfo{year}{2018}\natexlab{a}.
\newblock \showarticletitle{The Effect of Daylight versus Darkness on Driver
  Sleepiness: A Driving Simulator Study}.
\newblock \bibinfo{journal}{\emph{J. Sleep Res.}} \bibinfo{volume}{27},
  \bibinfo{number}{3} (\bibinfo{date}{June} \bibinfo{year}{2018}),
  \bibinfo{pages}{1--9}.
\newblock
\href{https://doi.org/10.1111/jsr.12642}{doi:\nolinkurl{10.1111/jsr.12642}}


\bibitem[Ahlstr{\"o}m et~al\mbox{.}(2018b)]%
        {Ahlstrom2018Effects}
\bibfield{author}{\bibinfo{person}{Christer Ahlstr{\"o}m},
  \bibinfo{person}{Anna Anund}, \bibinfo{person}{Carina Fors}, {and}
  \bibinfo{person}{Torbj{\"o}rn {\AA}kerstedt}.}
  \bibinfo{year}{2018}\natexlab{b}.
\newblock \showarticletitle{Effects of the Road Environment on the Development
  of Driver Sleepiness in Young Male Drivers}.
\newblock \bibinfo{journal}{\emph{Accid. Anal. \& Prev.}}
  \bibinfo{volume}{112} (\bibinfo{date}{March} \bibinfo{year}{2018}),
  \bibinfo{pages}{127--134}.
\newblock
\href{https://doi.org/10.1016/j.aap.2018.01.012}{doi:\nolinkurl{10.1016/j.aap.2018.01.012}}


\bibitem[Ahlstr{\"o}m et~al\mbox{.}(2021)]%
        {Ahlstrom2021Effects}
\bibfield{author}{\bibinfo{person}{Christer Ahlstr{\"o}m},
  \bibinfo{person}{Raimondas Zemblys}, \bibinfo{person}{Herman Jansson},
  \bibinfo{person}{Christian Forsberg}, \bibinfo{person}{Johan Karlsson}, {and}
  \bibinfo{person}{Anna Anund}.} \bibinfo{year}{2021}\natexlab{}.
\newblock \showarticletitle{Effects of Partially Automated Driving on the
  Development of Driver Sleepiness}.
\newblock \bibinfo{journal}{\emph{Accid. Anal. \& Prev.}}
  \bibinfo{volume}{153} (\bibinfo{date}{April} \bibinfo{year}{2021}),
  \bibinfo{pages}{1--9}.
\newblock
\href{https://doi.org/10.1016/j.aap.2021.106058}{doi:\nolinkurl{10.1016/j.aap.2021.106058}}


\bibitem[Bai and Feng(2025)]%
        {Bai2025Awakening}
\bibfield{author}{\bibinfo{person}{Xiaolu Bai} {and} \bibinfo{person}{Jing
  Feng}.} \bibinfo{year}{2025}\natexlab{}.
\newblock \showarticletitle{Awakening the Disengaged: {{Can}} Driving-Related
  Prompts Engage Drivers in Partial Automation?}
\newblock \bibinfo{journal}{\emph{Hum. Factors: J. Hum. Factors Ergon. Soc.}}
  \bibinfo{volume}{Online first} (\bibinfo{date}{Jan.} \bibinfo{year}{2025}),
  \bibinfo{pages}{1--22}.
\newblock
\href{https://doi.org/10.1177/00187208251314248}{doi:\nolinkurl{10.1177/00187208251314248}}


\bibitem[Beggiato and Krems(2013)]%
        {Beggiato2013TheEvolution}
\bibfield{author}{\bibinfo{person}{Matthias Beggiato} {and}
  \bibinfo{person}{Josef~F. Krems}.} \bibinfo{year}{2013}\natexlab{}.
\newblock \showarticletitle{The Evolution of Mental Model, Trust and Acceptance
  of Adaptive Cruise Control in Relation to Initial Information}.
\newblock \bibinfo{journal}{\emph{Transp. Res. Part F: Traffic Psychol.
  Behav.}}  \bibinfo{volume}{18} (\bibinfo{date}{May} \bibinfo{year}{2013}),
  \bibinfo{pages}{47--57}.
\newblock
\href{https://doi.org/10.1016/j.trf.2012.12.006}{doi:\nolinkurl{10.1016/j.trf.2012.12.006}}


\bibitem[Bianchi~Piccinini et~al\mbox{.}(2014)]%
        {BianchiPiccinini2014Drivers}
\bibfield{author}{\bibinfo{person}{Giulio~Francesco Bianchi~Piccinini},
  \bibinfo{person}{Carlos~Manuel Rodrigues}, \bibinfo{person}{Miguel
  Leit{\~a}o}, {and} \bibinfo{person}{Anabela Sim{\~o}es}.}
  \bibinfo{year}{2014}\natexlab{}.
\newblock \showarticletitle{Driver's Behavioral Adaptation to {{Adaptive Cruise
  Control}} ({{ACC}}): {{The}} Case of Speed and Time Headway}.
\newblock \bibinfo{journal}{\emph{J. Saf. Res.}}  \bibinfo{volume}{49}
  (\bibinfo{date}{June} \bibinfo{year}{2014}), \bibinfo{pages}{77--84}.
\newblock
\href{https://doi.org/10.1016/j.jsr.2014.02.010}{doi:\nolinkurl{10.1016/j.jsr.2014.02.010}}


\bibitem[Bl{\"o}macher et~al\mbox{.}(2020)]%
        {Blomacher2020TheEvolution}
\bibfield{author}{\bibinfo{person}{Katja Bl{\"o}macher},
  \bibinfo{person}{Gerhard N{\"o}cker}, {and} \bibinfo{person}{Markus Huff}.}
  \bibinfo{year}{2020}\natexlab{}.
\newblock \showarticletitle{The Evolution of Mental Models in Relation to
  Initial Information While Driving Automated}.
\newblock \bibinfo{journal}{\emph{Transp. Res. Part F: Traffic Psychol.
  Behav.}}  \bibinfo{volume}{68} (\bibinfo{date}{Jan.} \bibinfo{year}{2020}),
  \bibinfo{pages}{198--217}.
\newblock
\href{https://doi.org/10.1016/j.trf.2019.11.003}{doi:\nolinkurl{10.1016/j.trf.2019.11.003}}


\bibitem[Buchner et~al\mbox{.}(2024)]%
        {Buchner2024Exploring}
\bibfield{author}{\bibinfo{person}{Claudia Buchner}, \bibinfo{person}{Chantal
  Himmels}, \bibinfo{person}{Jan Schmitz}, \bibinfo{person}{Tanja Stoll}, {and}
  \bibinfo{person}{Martin Baumann}.} \bibinfo{year}{2024}\natexlab{}.
\newblock \showarticletitle{Exploring Urban Challenges: {{Understanding}}
  Advanced Driver Assistance Systems in Different Situational Contexts}. In
  \bibinfo{booktitle}{\emph{Int. Conf. Automot. User Interfaces Interact. Veh.
  Appl. (AutomotiveUI)}}. \bibinfo{publisher}{ACM}, \bibinfo{address}{Stanford,
  CA, USA}, \bibinfo{pages}{1--12}.
\newblock
\href{https://doi.org/10.1145/3640792.3675709}{doi:\nolinkurl{10.1145/3640792.3675709}}


\bibitem[Cabrall et~al\mbox{.}(2018)]%
        {Cabrall2018Adaptive}
\bibfield{author}{\bibinfo{person}{Christopher D.~D. Cabrall},
  \bibinfo{person}{Nico~M. Janssen}, {and} \bibinfo{person}{Joost C.~F. {de
  Winter}}.} \bibinfo{year}{2018}\natexlab{}.
\newblock \showarticletitle{Adaptive Automation: Automatically (Dis)Engaging
  Automation during Visually Distracted Driving}.
\newblock \bibinfo{journal}{\emph{PeerJ Computer Science}}  \bibinfo{volume}{4}
  (\bibinfo{date}{Oct.} \bibinfo{year}{2018}), \bibinfo{pages}{1--27}.
\newblock
\href{https://doi.org/10.7717/peerj-cs.166}{doi:\nolinkurl{10.7717/peerj-cs.166}}


\bibitem[Coughlin et~al\mbox{.}(2011)]%
        {Coughlin2011Monitoring}
\bibfield{author}{\bibinfo{person}{Joseph~F. Coughlin}, \bibinfo{person}{Bryan
  Reimer}, {and} \bibinfo{person}{Bruce Mehler}.}
  \bibinfo{year}{2011}\natexlab{}.
\newblock \showarticletitle{Monitoring, Managing, and Motivating Driver Safety
  and Well-Being}.
\newblock \bibinfo{journal}{\emph{IEEE Pervasive Comput.}}
  \bibinfo{volume}{10}, \bibinfo{number}{3} (\bibinfo{year}{2011}),
  \bibinfo{pages}{14--21}.
\newblock
\href{https://doi.org/10.1109/MPRV.2011.54}{doi:\nolinkurl{10.1109/MPRV.2011.54}}


\bibitem[{de Winter} et~al\mbox{.}(2014)]%
        {DeWinter2014Effects}
\bibfield{author}{\bibinfo{person}{Joost C.~F. {de Winter}},
  \bibinfo{person}{Riender Happee}, \bibinfo{person}{Marieke~H. Martens}, {and}
  \bibinfo{person}{Neville~A. Stanton}.} \bibinfo{year}{2014}\natexlab{}.
\newblock \showarticletitle{Effects of Adaptive Cruise Control and Highly
  Automated Driving on Workload and Situation Awareness: {{A}} Review of the
  Empirical Evidence}.
\newblock \bibinfo{journal}{\emph{Transp. Res. Part F: Traffic Psychol.
  Behav.}}  \bibinfo{volume}{27} (\bibinfo{date}{Nov.} \bibinfo{year}{2014}),
  \bibinfo{pages}{196--217}.
\newblock
\href{https://doi.org/10.1016/j.trf.2014.06.016}{doi:\nolinkurl{10.1016/j.trf.2014.06.016}}


\bibitem[Dixon(2020)]%
        {Dixon2020Autonowashing}
\bibfield{author}{\bibinfo{person}{Liza Dixon}.}
  \bibinfo{year}{2020}\natexlab{}.
\newblock \showarticletitle{Autonowashing: {{The}} Greenwashing of Vehicle
  Automation}.
\newblock \bibinfo{journal}{\emph{Transp. Res. Interdiscip. Perspect.}}
  \bibinfo{volume}{5} (\bibinfo{date}{May} \bibinfo{year}{2020}),
  \bibinfo{pages}{1--9}.
\newblock
\href{https://doi.org/10.1016/j.trip.2020.100113}{doi:\nolinkurl{10.1016/j.trip.2020.100113}}


\bibitem[Dixon et~al\mbox{.}(2025)]%
        {Dixon2025Explaining}
\bibfield{author}{\bibinfo{person}{Liza Dixon}, \bibinfo{person}{Norbert
  Schneider}, \bibinfo{person}{Dominik M{\"u}hlbacher}, \bibinfo{person}{Dennis
  Befelein}, \bibinfo{person}{Frank~O. Flemisch}, {and} \bibinfo{person}{Martin
  Baumann}.} \bibinfo{year}{2025}\natexlab{}.
\newblock \showarticletitle{Explaining Authoritative Control Interventions in
  Automated Driving to Support Driver Understanding, Trust, and Reliance}.
\newblock \bibinfo{journal}{\emph{Transp. Res. Part F: Traffic Psychol.
  Behav.}}  \bibinfo{volume}{113} (\bibinfo{date}{Aug.} \bibinfo{year}{2025}),
  \bibinfo{pages}{194--212}.
\newblock
\href{https://doi.org/10.1016/j.trf.2025.04.013}{doi:\nolinkurl{10.1016/j.trf.2025.04.013}}


\bibitem[Dosovitskiy et~al\mbox{.}(2017)]%
        {Dosovitskiy2017CARLA}
\bibfield{author}{\bibinfo{person}{Alexey Dosovitskiy}, \bibinfo{person}{German
  Ros}, \bibinfo{person}{Felipe Codevilla}, \bibinfo{person}{Antonio Lopez},
  {and} \bibinfo{person}{Vladlen Koltun}.} \bibinfo{year}{2017}\natexlab{}.
\newblock \showarticletitle{{{CARLA}}: {{An}} Open Urban Driving Simulator}. In
  \bibinfo{booktitle}{\emph{Annu. Conf. Robot. Learn.}}
  \emph{(\bibinfo{series}{Proc. Mach. Learn. Res.},
  Vol.~\bibinfo{volume}{78})}. \bibinfo{publisher}{ML Research Press},
  \bibinfo{address}{Mountain View, CA, USA}, \bibinfo{pages}{1--16}.
\newblock


\bibitem[Du et~al\mbox{.}(2020)]%
        {Du2020Evaluating}
\bibfield{author}{\bibinfo{person}{Na Du}, \bibinfo{person}{Jinyong Kim},
  \bibinfo{person}{Feng Zhou}, \bibinfo{person}{Elizabeth Pulver},
  \bibinfo{person}{Dawn~M. Tilbury}, \bibinfo{person}{Lionel~P. Robert},
  \bibinfo{person}{Anuj~K. Pradhan}, {and} \bibinfo{person}{X.~Jessie Yang}.}
  \bibinfo{year}{2020}\natexlab{}.
\newblock \showarticletitle{Evaluating Effects of Cognitive Load, Takeover
  Request Lead Time, and Traffic Density on Drivers' Takeover Performance in
  Conditionally Automated Driving}. In \bibinfo{booktitle}{\emph{Int. Conf.
  Automot. User Interfaces Interact. Veh. Appl. (AutomotiveUI)}}.
  \bibinfo{publisher}{ACM}, \bibinfo{address}{Virtual Conf.},
  \bibinfo{pages}{66--73}.
\newblock
\href{https://doi.org/10.1145/3409120.3410666}{doi:\nolinkurl{10.1145/3409120.3410666}}


\bibitem[Endsley(2016)]%
        {Endsley2016From}
\bibfield{author}{\bibinfo{person}{Mica~R. Endsley}.}
  \bibinfo{year}{2016}\natexlab{}.
\newblock \showarticletitle{From Here to Autonomy}.
\newblock \bibinfo{journal}{\emph{Hum. Factors: J. Hum. Factors Ergon. Soc.}}
  \bibinfo{volume}{59}, \bibinfo{number}{1} (\bibinfo{date}{Dec.}
  \bibinfo{year}{2016}), \bibinfo{pages}{5--27}.
\newblock
\href{https://doi.org/10.1177/0018720816681350}{doi:\nolinkurl{10.1177/0018720816681350}}


\bibitem[Gershon et~al\mbox{.}(2021)]%
        {Gershon2021Driver}
\bibfield{author}{\bibinfo{person}{Pnina Gershon}, \bibinfo{person}{Sean
  Seaman}, \bibinfo{person}{Bruce Mehler}, \bibinfo{person}{Bryan Reimer},
  {and} \bibinfo{person}{Joseph Coughlin}.} \bibinfo{year}{2021}\natexlab{}.
\newblock \showarticletitle{Driver Behavior and the Use of Automation in
  Real-World Driving}.
\newblock \bibinfo{journal}{\emph{Accid. Anal. \& Prev.}}
  \bibinfo{volume}{158} (\bibinfo{date}{Aug.} \bibinfo{year}{2021}),
  \bibinfo{pages}{1--9}.
\newblock
\href{https://doi.org/10.1016/j.aap.2021.106217}{doi:\nolinkurl{10.1016/j.aap.2021.106217}}


\bibitem[Gold et~al\mbox{.}(2016)]%
        {Gold2016Taking}
\bibfield{author}{\bibinfo{person}{Christian Gold}, \bibinfo{person}{Moritz
  K{\"o}rber}, \bibinfo{person}{David Lechner}, {and} \bibinfo{person}{Klaus
  Bengler}.} \bibinfo{year}{2016}\natexlab{}.
\newblock \showarticletitle{Taking over Control from Highly Automated Vehicles
  in Complex Traffic Situations}.
\newblock \bibinfo{journal}{\emph{Hum. Factors: J. Hum. Factors Ergon. Soc.}}
  \bibinfo{volume}{58}, \bibinfo{number}{4} (\bibinfo{date}{March}
  \bibinfo{year}{2016}), \bibinfo{pages}{642--652}.
\newblock
\href{https://doi.org/10.1177/0018720816634226}{doi:\nolinkurl{10.1177/0018720816634226}}


\bibitem[Griffiths and Gillespie(2005)]%
        {Griffiths2005Sharing}
\bibfield{author}{\bibinfo{person}{Paul~G. Griffiths} {and}
  \bibinfo{person}{R.~Brent Gillespie}.} \bibinfo{year}{2005}\natexlab{}.
\newblock \showarticletitle{Sharing Control between Humans and Automation Using
  Haptic Interface: {{Primary}} and Secondary Task Performance Benefits}.
\newblock \bibinfo{journal}{\emph{Hum. Factors: J. Hum. Factors Ergon. Soc.}}
  \bibinfo{volume}{47}, \bibinfo{number}{3} (\bibinfo{date}{Sept.}
  \bibinfo{year}{2005}), \bibinfo{pages}{574--590}.
\newblock
\href{https://doi.org/10.1518/001872005774859944}{doi:\nolinkurl{10.1518/001872005774859944}}


\bibitem[Halin et~al\mbox{.}(2021)]%
        {Halin2021Survey}
\bibfield{author}{\bibinfo{person}{Ana{\"{\i}}s Halin},
  \bibinfo{person}{Jacques~G. Verly}, {and} \bibinfo{person}{Marc
  Van~Droogenbroeck}.} \bibinfo{year}{2021}\natexlab{}.
\newblock \showarticletitle{Survey and Synthesis of State of the Art in Driver
  Monitoring}.
\newblock \bibinfo{journal}{\emph{Sensors}} \bibinfo{volume}{21},
  \bibinfo{number}{16} (\bibinfo{date}{Aug.} \bibinfo{year}{2021}),
  \bibinfo{pages}{1--48}.
\newblock
\href{https://doi.org/10.3390/s21165558}{doi:\nolinkurl{10.3390/s21165558}}


\bibitem[Herzberger et~al\mbox{.}(2022)]%
        {Herzberger2022Confidence}
\bibfield{author}{\bibinfo{person}{Nicolas Herzberger}, \bibinfo{person}{Marcel
  Usai}, {and} \bibinfo{person}{Frank Flemisch}.}
  \bibinfo{year}{2022}\natexlab{}.
\newblock \showarticletitle{Confidence Horizon for a Dynamic Balance between
  Drivers and Vehicle Automation: {{First}} Sketch and Application}.
\newblock \bibinfo{journal}{\emph{Hum. Factors Transp.}}  \bibinfo{volume}{60}
  (\bibinfo{year}{2022}), \bibinfo{pages}{33--39}.
\newblock
\href{https://doi.org/10.54941/ahfe1002431}{doi:\nolinkurl{10.54941/ahfe1002431}}


\bibitem[Herzberger et~al\mbox{.}(2024)]%
        {Herzberger2024Cooperation}
\bibfield{author}{\bibinfo{person}{Nicolas Herzberger}, \bibinfo{person}{Marcel
  Usai}, \bibinfo{person}{Maximilian Schwalm}, {and} \bibinfo{person}{Frank
  Flemisch}.} \bibinfo{year}{2024}\natexlab{}.
\newblock \showarticletitle{Cooperation between Vehicle and Driver:
  {{Predicting}} the Driver's Takeover Capability in Cooperative Automated
  Driving Based on Orientation Patterns}.
\newblock In \bibinfo{booktitle}{\emph{Cooperatively Interacting Vehicles}}.
  \bibinfo{publisher}{Springer Int. Publ.}, \bibinfo{address}{Cham, Switz.},
  \bibinfo{pages}{509--523}.
\newblock
\href{https://doi.org/10.1007/978-3-031-60494-2_17}{doi:\nolinkurl{10.1007/978-3-031-60494-2_17}}


\bibitem[Hoedemaeker and Brookhuis(1998)]%
        {Hoedemaeker1998Behavioural}
\bibfield{author}{\bibinfo{person}{Marika Hoedemaeker} {and}
  \bibinfo{person}{Karel~A. Brookhuis}.} \bibinfo{year}{1998}\natexlab{}.
\newblock \showarticletitle{Behavioural Adaptation to Driving with an Adaptive
  Cruise Control ({{ACC}})}.
\newblock \bibinfo{journal}{\emph{Transp. Res. Part F: Traffic Psychol.
  Behav.}} \bibinfo{volume}{1}, \bibinfo{number}{2} (\bibinfo{date}{Dec.}
  \bibinfo{year}{1998}), \bibinfo{pages}{95--106}.
\newblock
\href{https://doi.org/10.1016/s1369-8478(98)00008-4}{doi:\nolinkurl{10.1016/s1369-8478(98)00008-4}}


\bibitem[{JASP Team}(2025)]%
        {JASP2025}
\bibfield{author}{\bibinfo{person}{{JASP Team}}.}
  \bibinfo{year}{2025}\natexlab{}.
\newblock \bibinfo{booktitle}{\emph{{JASP} ({Version} 0.19.3): {Computer}
  Software}}.
\newblock Amsterdam, The Netherlands.
\newblock
\urldef\tempurl%
\url{https://jasp-stats.org}
\showURL{%
\tempurl}


\bibitem[Li et~al\mbox{.}(2021)]%
        {Li2021Comprehensive}
\bibfield{author}{\bibinfo{person}{Yang Li}, \bibinfo{person}{Linbo Li},
  \bibinfo{person}{Daiheng Ni}, {and} \bibinfo{person}{Yue Zhang}.}
  \bibinfo{year}{2021}\natexlab{}.
\newblock \showarticletitle{Comprehensive Survival Analysis of Lane-Changing
  Duration}.
\newblock \bibinfo{journal}{\emph{Measurement}}  \bibinfo{volume}{182}
  (\bibinfo{date}{Sept.} \bibinfo{year}{2021}), \bibinfo{pages}{1--15}.
\newblock
\href{https://doi.org/10.1016/j.measurement.2021.109707}{doi:\nolinkurl{10.1016/j.measurement.2021.109707}}


\bibitem[Liang and Lee(2010)]%
        {Liang2010Combining}
\bibfield{author}{\bibinfo{person}{Yulan Liang} {and} \bibinfo{person}{John~D.
  Lee}.} \bibinfo{year}{2010}\natexlab{}.
\newblock \showarticletitle{Combining Cognitive and Visual Distraction:
  {{Less}} than the Sum of Its Parts}.
\newblock \bibinfo{journal}{\emph{Accid. Anal. \& Prev.}} \bibinfo{volume}{42},
  \bibinfo{number}{3} (\bibinfo{date}{May} \bibinfo{year}{2010}),
  \bibinfo{pages}{881--890}.
\newblock
\href{https://doi.org/10.1016/j.aap.2009.05.001}{doi:\nolinkurl{10.1016/j.aap.2009.05.001}}


\bibitem[Lin and Chen(2024)]%
        {Lin2024How}
\bibfield{author}{\bibinfo{person}{Zijian Lin} {and} \bibinfo{person}{Feng
  Chen}.} \bibinfo{year}{2024}\natexlab{}.
\newblock \showarticletitle{How Various Urgencies and Visibilities Influence
  Drivers' Takeover Performance in Critical Car-Following Conditions? {{A}}
  Driving Simulation Study}.
\newblock \bibinfo{journal}{\emph{Transp. Res. Part F: Traffic Psychol.
  Behav.}}  \bibinfo{volume}{104} (\bibinfo{date}{July} \bibinfo{year}{2024}),
  \bibinfo{pages}{303--317}.
\newblock
\href{https://doi.org/10.1016/j.trf.2024.06.007}{doi:\nolinkurl{10.1016/j.trf.2024.06.007}}


\bibitem[Meyerson et~al\mbox{.}(2024)]%
        {Meyerson2024Effects}
\bibfield{author}{\bibinfo{person}{Amanda Meyerson}, \bibinfo{person}{Johanna
  Eklind}, \bibinfo{person}{Florian Fischer}, \bibinfo{person}{Maytheewat
  Aramrattana}, \bibinfo{person}{Ingemar Fredriksson}, {and}
  \bibinfo{person}{Christer Ahlstr{\"o}m}.} \bibinfo{year}{2024}\natexlab{}.
\newblock \showarticletitle{Effects of Daylight and Darkness at Daytime versus
  Nighttime on Driver Sleepiness: {{A}} Driving Simulator Study}.
\newblock \bibinfo{journal}{\emph{Transp. Res. Interdiscip. Perspect.}}
  \bibinfo{volume}{24} (\bibinfo{date}{March} \bibinfo{year}{2024}),
  \bibinfo{pages}{101087}.
\newblock
\href{https://doi.org/10.1016/j.trip.2024.101087}{doi:\nolinkurl{10.1016/j.trip.2024.101087}}


\bibitem[Micaelli and Samson(1993)]%
        {Micaelli1993Trajectory}
\bibfield{author}{\bibinfo{person}{Alain Micaelli} {and}
  \bibinfo{person}{Claude Samson}.} \bibinfo{year}{1993}\natexlab{}.
\newblock \bibinfo{booktitle}{\emph{Trajectory Tracking for Unicycle-Type and
  Two-Steering-Wheels Mobile Robots}}.
\newblock \bibinfo{type}{{T}echnical {R}eport}. \bibinfo{institution}{INRIA}.
\newblock


\bibitem[Naujoks et~al\mbox{.}(2016)]%
        {Naujoks2016AHumanMachine}
\bibfield{author}{\bibinfo{person}{Frederik Naujoks}, \bibinfo{person}{Yannick
  Forster}, \bibinfo{person}{Katharina Wiedemann}, {and}
  \bibinfo{person}{Alexandra Neukum}.} \bibinfo{year}{2016}\natexlab{}.
\newblock \showarticletitle{A Human-Machine Interface for Cooperative Highly
  Automated Driving}. In \bibinfo{booktitle}{\emph{Advances in Human Aspects of
  Transportation}} \emph{(\bibinfo{series}{Adv. Intell. Syst. Comput.},
  Vol.~\bibinfo{volume}{484})}. \bibinfo{publisher}{Springer Int. Publ.},
  \bibinfo{address}{Orlando, Florida, USA}, \bibinfo{pages}{585--595}.
\newblock
\href{https://doi.org/10.1007/978-3-319-41682-3_49}{doi:\nolinkurl{10.1007/978-3-319-41682-3_49}}


\bibitem[Novakazi et~al\mbox{.}(2021)]%
        {Novakazi2021Levels}
\bibfield{author}{\bibinfo{person}{Fjoll{\"e} Novakazi},
  \bibinfo{person}{Mikael Johansson}, \bibinfo{person}{Helena Str{\"o}mberg},
  {and} \bibinfo{person}{MariAnne Karlsson}.} \bibinfo{year}{2021}\natexlab{}.
\newblock \showarticletitle{Levels of What? {{Investigating}} Drivers'
  Understanding of Different Levels of Automation in Vehicles}.
\newblock \bibinfo{journal}{\emph{J. Cogn. Eng. Decis. Mak.}}
  \bibinfo{volume}{15}, \bibinfo{number}{2-3} (\bibinfo{date}{April}
  \bibinfo{year}{2021}), \bibinfo{pages}{116--132}.
\newblock
\href{https://doi.org/10.1177/15553434211009024}{doi:\nolinkurl{10.1177/15553434211009024}}


\bibitem[Orlovska et~al\mbox{.}(2020)]%
        {Orlovska2020Effects}
\bibfield{author}{\bibinfo{person}{Julia Orlovska}, \bibinfo{person}{Fjoll{\"e}
  Novakazi}, \bibinfo{person}{Blig{\aa}rd {Lars-Ola}},
  \bibinfo{person}{MariAnne Karlsson}, \bibinfo{person}{Casper Wickman}, {and}
  \bibinfo{person}{Rikard S{\"o}derberg}.} \bibinfo{year}{2020}\natexlab{}.
\newblock \showarticletitle{Effects of the Driving Context on the Usage of
  Automated Driver Assistance Systems ({{ADAS}}) -- Naturalistic Driving Study
  for {{ADAS}} Evaluation}.
\newblock \bibinfo{journal}{\emph{Transp. Res. Interdiscip. Perspect.}}
  \bibinfo{volume}{4} (\bibinfo{date}{March} \bibinfo{year}{2020}),
  \bibinfo{pages}{1--16}.
\newblock
\href{https://doi.org/10.1016/j.trip.2020.100093}{doi:\nolinkurl{10.1016/j.trip.2020.100093}}


\bibitem[Park et~al\mbox{.}(2022)]%
        {Park2022TheImpact}
\bibfield{author}{\bibinfo{person}{Sami Park}, \bibinfo{person}{Yilun Xing},
  \bibinfo{person}{Kumar Akash}, \bibinfo{person}{Teruhisa Misu}, {and}
  \bibinfo{person}{Linda~Ng Boyle}.} \bibinfo{year}{2022}\natexlab{}.
\newblock \showarticletitle{The Impact of Environmental Complexity on Drivers'
  Situation Awareness}. In \bibinfo{booktitle}{\emph{Int. Conf. Automot. User
  Interfaces Interact. Veh. Appl. (AutomotiveUI)}}. \bibinfo{publisher}{ACM},
  \bibinfo{address}{Seoul, South Korea}, \bibinfo{pages}{131--138}.
\newblock
\href{https://doi.org/10.1145/3543174.3546831}{doi:\nolinkurl{10.1145/3543174.3546831}}


\bibitem[Radlmayr et~al\mbox{.}(2014)]%
        {Radlmayr2014How}
\bibfield{author}{\bibinfo{person}{Jonas Radlmayr}, \bibinfo{person}{Christian
  Gold}, \bibinfo{person}{Lutz Lorenz}, \bibinfo{person}{Mehdi Farid}, {and}
  \bibinfo{person}{Klaus Bengler}.} \bibinfo{year}{2014}\natexlab{}.
\newblock \showarticletitle{How Traffic Situations and Non-Driving Related
  Tasks Affect the Take-over Quality in Highly Automated Driving}.
\newblock \bibinfo{journal}{\emph{Proc. Hum. Factors Ergon. Soc. Annu. Meet.}}
  \bibinfo{volume}{58}, \bibinfo{number}{1} (\bibinfo{date}{Sept.}
  \bibinfo{year}{2014}), \bibinfo{pages}{2063--2067}.
\newblock
\href{https://doi.org/10.1177/1541931214581434}{doi:\nolinkurl{10.1177/1541931214581434}}


\bibitem[Ranney(2008)]%
        {Ranney2008Driver}
\bibfield{author}{\bibinfo{person}{T. Ranney}.}
  \bibinfo{year}{2008}\natexlab{}.
\newblock \bibinfo{booktitle}{\emph{Driver Distraction: {A} Review of the
  Current State-of-Knowledge}}.
\newblock \bibinfo{type}{{T}echnical {R}eport}. \bibinfo{institution}{National
  Highw. Traffic Saf. Adm.}
\newblock


\bibitem[Richardson(2011)]%
        {Richardson2011Eta}
\bibfield{author}{\bibinfo{person}{John T.~E. Richardson}.}
  \bibinfo{year}{2011}\natexlab{}.
\newblock \showarticletitle{Eta Squared and Partial Eta Squared as Measures of
  Effect Size in Educational Research}.
\newblock \bibinfo{journal}{\emph{Educ. Res. Rev.}} \bibinfo{volume}{6},
  \bibinfo{number}{2} (\bibinfo{date}{Jan.} \bibinfo{year}{2011}),
  \bibinfo{pages}{135--147}.
\newblock
\href{https://doi.org/10.1016/j.edurev.2010.12.001}{doi:\nolinkurl{10.1016/j.edurev.2010.12.001}}


\bibitem[{SAE International}(2021)]%
        {Sae2021Taxonomy}
\bibfield{author}{\bibinfo{person}{{SAE International}}.}
  \bibinfo{year}{2021}\natexlab{}.
\newblock \bibinfo{booktitle}{\emph{Taxonomy and Definitions for Terms Related
  to Driving Automation Systems for On-Road Motor Vehicles}}.
\newblock \bibinfo{type}{{T}echnical {R}eport} SAE Standard J3016 202104.
  \bibinfo{institution}{Society of Automobile Engineers},
  \bibinfo{address}{Warrendale, PA, USA}. \bibinfo{pages}{1--41} pages.
\newblock
\href{https://doi.org/10.4271/J3016_202104}{doi:\nolinkurl{10.4271/J3016_202104}}


\bibitem[Sever and Contissa(2024)]%
        {Sever2024Automated}
\bibfield{author}{\bibinfo{person}{Tina Sever} {and} \bibinfo{person}{Giuseppe
  Contissa}.} \bibinfo{year}{2024}\natexlab{}.
\newblock \showarticletitle{Automated Driving Regulations - Where Are We Now?}
\newblock \bibinfo{journal}{\emph{Transp. Res. Interdiscip. Perspect.}}
  \bibinfo{volume}{24} (\bibinfo{date}{March} \bibinfo{year}{2024}),
  \bibinfo{pages}{101033}.
\newblock
\href{https://doi.org/10.1016/j.trip.2024.101033}{doi:\nolinkurl{10.1016/j.trip.2024.101033}}


\bibitem[Stanton and Young(2005)]%
        {Stanton2005Driver}
\bibfield{author}{\bibinfo{person}{Neville~A. Stanton} {and}
  \bibinfo{person}{Mark~S. Young}.} \bibinfo{year}{2005}\natexlab{}.
\newblock \showarticletitle{Driver Behaviour with Adaptive Cruise Control}.
\newblock \bibinfo{journal}{\emph{Ergonomics}} \bibinfo{volume}{48},
  \bibinfo{number}{10} (\bibinfo{date}{Aug.} \bibinfo{year}{2005}),
  \bibinfo{pages}{1294--1313}.
\newblock
\href{https://doi.org/10.1080/00140130500252990}{doi:\nolinkurl{10.1080/00140130500252990}}


\bibitem[Sujit et~al\mbox{.}(2014)]%
        {Sujit2014Unmanned}
\bibfield{author}{\bibinfo{person}{P.~B. Sujit}, \bibinfo{person}{Srikanth
  Saripalli}, {and} \bibinfo{person}{Joao~Borges Sousa}.}
  \bibinfo{year}{2014}\natexlab{}.
\newblock \showarticletitle{Unmanned Aerial Vehicle Path Following: A Survey
  and Analysis of Algorithms for Fixed-Wing Unmanned Aerial Vehicles}.
\newblock \bibinfo{journal}{\emph{IEEE Control Syst. Mag.}}
  \bibinfo{volume}{34}, \bibinfo{number}{1} (\bibinfo{date}{Feb.}
  \bibinfo{year}{2014}), \bibinfo{pages}{42--59}.
\newblock
\href{https://doi.org/10.1109/MCS.2013.2287568}{doi:\nolinkurl{10.1109/MCS.2013.2287568}}


\bibitem[Vasta and Biondi(2025)]%
        {Vasta2025Effect}
\bibfield{author}{\bibinfo{person}{Nicola Vasta} {and}
  \bibinfo{person}{Francesco Biondi}.} \bibinfo{year}{2025}\natexlab{}.
\newblock \showarticletitle{Effect of Partially Automated Driving on Mental
  Workload, Visual Behavior and Engagement in Nondriving-Related Tasks: A
  Meta-Analysis}.
\newblock \bibinfo{journal}{\emph{Hum. Factors: J. Hum. Factors Ergon. Soc.}}
  \bibinfo{volume}{Online first} (\bibinfo{date}{March} \bibinfo{year}{2025}),
  \bibinfo{pages}{1--30}.
\newblock
\href{https://doi.org/10.1177/00187208251323132}{doi:\nolinkurl{10.1177/00187208251323132}}


\bibitem[Vogelpohl et~al\mbox{.}(2019)]%
        {Vogelpohl2019Asleep}
\bibfield{author}{\bibinfo{person}{Tobias Vogelpohl}, \bibinfo{person}{Matthias
  K{\"u}hn}, \bibinfo{person}{Thomas Hummel}, {and} \bibinfo{person}{Mark
  Vollrath}.} \bibinfo{year}{2019}\natexlab{}.
\newblock \showarticletitle{Asleep at the Automated Wheel --- {{Sleepiness}}
  and Fatigue during Highly Automated Driving}.
\newblock \bibinfo{journal}{\emph{Accid. Anal. \& Prev.}}
  \bibinfo{volume}{126} (\bibinfo{date}{May} \bibinfo{year}{2019}),
  \bibinfo{pages}{70--84}.
\newblock
\href{https://doi.org/10.1016/j.aap.2018.03.013}{doi:\nolinkurl{10.1016/j.aap.2018.03.013}}


\bibitem[Vollrath et~al\mbox{.}(2011)]%
        {Vollrath2011TheInfluence}
\bibfield{author}{\bibinfo{person}{Mark Vollrath}, \bibinfo{person}{Susanne
  Schleicher}, {and} \bibinfo{person}{Christhard Gelau}.}
  \bibinfo{year}{2011}\natexlab{}.
\newblock \showarticletitle{The Influence of {{Cruise Control}} and {{Adaptive
  Cruise Control}} on Driving Behaviour -- {{A}} Driving Simulator Study}.
\newblock \bibinfo{journal}{\emph{Accid. Anal. \& Prev.}} \bibinfo{volume}{43},
  \bibinfo{number}{3} (\bibinfo{date}{May} \bibinfo{year}{2011}),
  \bibinfo{pages}{1134--1139}.
\newblock
\href{https://doi.org/10.1016/j.aap.2010.12.023}{doi:\nolinkurl{10.1016/j.aap.2010.12.023}}


\bibitem[Walch et~al\mbox{.}(2019)]%
        {Walch2019Driving}
\bibfield{author}{\bibinfo{person}{Marcel Walch}, \bibinfo{person}{Mark
  Colley}, {and} \bibinfo{person}{Michael Weber}.}
  \bibinfo{year}{2019}\natexlab{}.
\newblock \showarticletitle{Driving-Task-Related Human-Machine Interaction in
  Automated Driving}. In \bibinfo{booktitle}{\emph{Adjun. Proc. Int. Conf.
  Automot. User Interfaces Interact. Veh. Appl. (AutomotiveUI)}}.
  \bibinfo{publisher}{ACM}, \bibinfo{address}{Utrecht, The Netherlands},
  \bibinfo{pages}{427--433}.
\newblock
\href{https://doi.org/10.1145/3349263.3351527}{doi:\nolinkurl{10.1145/3349263.3351527}}


\bibitem[Watzenig and Horn(2017)]%
        {Watzenig2017Automated}
\bibfield{author}{\bibinfo{person}{Daniel Watzenig} {and}
  \bibinfo{person}{Martin Horn}.} \bibinfo{year}{2017}\natexlab{}.
\newblock \bibinfo{booktitle}{\emph{Automated Driving: {{Safer}} and More
  Efficient Future Driving}}.
\newblock \bibinfo{publisher}{Springer}, \bibinfo{address}{Cham, Switz.}
\newblock
\href{https://doi.org/10.1007/978-3-319-31895-0}{doi:\nolinkurl{10.1007/978-3-319-31895-0}}


\bibitem[Wu and Boyle(2015)]%
        {Wu2015Drivers}
\bibfield{author}{\bibinfo{person}{Yuqing Wu} {and} \bibinfo{person}{Linda~Ng
  Boyle}.} \bibinfo{year}{2015}\natexlab{}.
\newblock \showarticletitle{Drivers' Engagement Level in {{Adaptive Cruise
  Control}} While Distracted or Impaired}.
\newblock \bibinfo{journal}{\emph{Transp. Res. Part F: Traffic Psychol.
  Behav.}}  \bibinfo{volume}{33} (\bibinfo{date}{Aug.} \bibinfo{year}{2015}),
  \bibinfo{pages}{7--15}.
\newblock
\href{https://doi.org/10.1016/j.trf.2015.05.005}{doi:\nolinkurl{10.1016/j.trf.2015.05.005}}


\bibitem[Xing et~al\mbox{.}(2021)]%
        {Xing2021Toward}
\bibfield{author}{\bibinfo{person}{Yang Xing}, \bibinfo{person}{Chen Lv},
  \bibinfo{person}{Dongpu Cao}, {and} \bibinfo{person}{Peng Hang}.}
  \bibinfo{year}{2021}\natexlab{}.
\newblock \showarticletitle{Toward Human-Vehicle Collaboration: {{Review}} and
  Perspectives on Human-Centered Collaborative Automated Driving}.
\newblock \bibinfo{journal}{\emph{Transp. Res. Part C: Emerg. Technol.}}
  \bibinfo{volume}{128} (\bibinfo{date}{July} \bibinfo{year}{2021}),
  \bibinfo{pages}{103199}.
\newblock
\href{https://doi.org/10.1016/j.trc.2021.103199}{doi:\nolinkurl{10.1016/j.trc.2021.103199}}


\bibitem[Yerkes and Dodson(1908)]%
        {Yerkes1908Relation}
\bibfield{author}{\bibinfo{person}{Robert~M. Yerkes} {and}
  \bibinfo{person}{John~D. Dodson}.} \bibinfo{year}{1908}\natexlab{}.
\newblock \showarticletitle{The Relation of Strength of Stimulus to Rapidity of
  Habit-formation}.
\newblock \bibinfo{journal}{\emph{J. Comp. Neurol. Psychol.}}
  \bibinfo{volume}{18}, \bibinfo{number}{5} (\bibinfo{date}{Nov.}
  \bibinfo{year}{1908}), \bibinfo{pages}{459--482}.
\newblock
\href{https://doi.org/10.1002/cne.920180503}{doi:\nolinkurl{10.1002/cne.920180503}}


\bibitem[Young and Stanton(2002)]%
        {Young2002Attention}
\bibfield{author}{\bibinfo{person}{Mark~S. Young} {and}
  \bibinfo{person}{Neville~A. Stanton}.} \bibinfo{year}{2002}\natexlab{}.
\newblock \showarticletitle{Attention and Automation: {{New}} Perspectives on
  Mental Underload and Performance}.
\newblock \bibinfo{journal}{\emph{Theor. Issues Ergon. Sci.}}
  \bibinfo{volume}{3}, \bibinfo{number}{2} (\bibinfo{date}{Jan.}
  \bibinfo{year}{2002}), \bibinfo{pages}{178--194}.
\newblock
\href{https://doi.org/10.1080/14639220210123789}{doi:\nolinkurl{10.1080/14639220210123789}}


\bibitem[Young and Stanton(2023)]%
        {Young2023ToAutomate}
\bibfield{author}{\bibinfo{person}{Mark~S. Young} {and}
  \bibinfo{person}{Neville~A. Stanton}.} \bibinfo{year}{2023}\natexlab{}.
\newblock \showarticletitle{To Automate or Not to Automate: Advocating the
  `cliff-Edge' Principle}.
\newblock \bibinfo{journal}{\emph{Ergonomics}} \bibinfo{volume}{66},
  \bibinfo{number}{11} (\bibinfo{date}{Oct.} \bibinfo{year}{2023}),
  \bibinfo{pages}{1695--1701}.
\newblock
\href{https://doi.org/10.1080/00140139.2023.2270786}{doi:\nolinkurl{10.1080/00140139.2023.2270786}}


\end{thebibliography}

\end{document}